\documentclass[journal]{IEEEtran}

\usepackage[cmex10]{amsmath}
\usepackage{amsfonts}
\usepackage{mathtools}
\usepackage{siunitx}
\usepackage{fixltx2e}
\usepackage{algorithm}
\usepackage{algorithmicx}
\usepackage[noend]{algpseudocode}
\usepackage[absolute]{textpos}
\usepackage[hyperindex,hidelinks]{hyperref}
\usepackage[hyphenbreaks]{breakurl}

\allowdisplaybreaks
\newcommand*\Let[2]{\State #1 $\gets$ #2}
\algrenewcommand\alglinenumber[1]{
    {\sf\footnotesize\addfontfeatures{Colour=888888,Numbers=Monospaced}#1}}
\algrenewcommand\algorithmicrequire{\textbf{Precondition:}}
\algrenewcommand\algorithmicensure{\textbf{Postcondition:}}
\MakeRobust{\Call}

\TPMargin{3pt}

\newcommand{\copyrightstatement}{ 
\begin{textblock}{0.84}(0.08,0.935) 
	\footnotesize
	\noindent
	\copyright~2015 IEEE. Personal use of this material is permitted. Permission from IEEE must be obtained for all other uses, in any current or future media, including reprinting/republishing this material for advertising or promotional purposes, creating new collective works, for resale or redistribution to servers or lists, or reuse of any copyrighted component of this work in other works. DOI: \href{http://dx.doi.org/10.1109/TNSM.2015.2464071}{10.1109/TNSM.2015.2464071}
\end{textblock}
}

\newcommand{\fref}[1]{Fig.~\ref{#1}}
\newcommand{\sref}[1]{Sec.~\ref{#1}}

\newcommand{\aref}[1]{Alg.~\ref{#1}}
\newcommand{\eref}[1]{(\ref{#1})}

\newcommand{\totalNumberOfVisits}{\ensuremath{K}}
\newcommand{\lifetimeFunctionSlope}{\ensuremath{m}}
\newcommand{\bigO}[1]{\ensuremath{\mathcal{O}(#1)}}
\newcommand{\maximumNumberOfVisits}{\ensuremath{\hat{K}}}

\newcommand{\sensorLocationSet}[1]{\ensuremath{\mathcal{T}_{#1}}}
\newcommand{\sinkLocationSet}[1]{\ensuremath{\mathcal{S}_{#1}}}
\newcommand{\candidateRelayLocationSet}[1]{\ensuremath{\mathcal{N}_{#1}}}
\newcommand{\universalSet}[1]{\ensuremath{\mathcal{U}_{#1}}}

\newcommand{\dataGenerationRate}[1]{\ensuremath{g_{#1}}}
\newcommand{\sensingPower}[1]{\ensuremath{S_{#1}}}
\newcommand{\distance}[2]{\ensuremath{d_{#1 #2}}}
\newcommand{\transmitPower}[1]{\ensuremath{T_{#1}(\distance{}{})}}
\newcommand{\receivePower}[1]{\ensuremath{Q_{#1}}}
\newcommand{\operationalLifetime}{\ensuremath{ L }}

\newcommand{\visitExpenditure}[1]{\ensuremath{p_{#1}}}
\newcommand{\visitNumber}{\ensuremath{k}}
\newcommand{\visitLifetime}[1]{\ensuremath{l_{#1}}}
\newcommand{\visitExpenditureIndex}{\ensuremath{n}}
\newcommand{\visitLifetimeFunction}[2]{\ensuremath{f_{#1}(#2)}}
\newcommand{\interestRate}{\ensuremath{ v }}

\newcommand{\unscheduledPayment}[1]{\ensuremath{\sigma_{#1}}}
\newcommand{\unscheduledPaymentTime}[1]{\ensuremath{t_{#1}}}
\newcommand{\numberOfFailures}{\ensuremath{F}}
\newcommand{\unscheduledPaymentNumber}{\ensuremath{n}}
\newcommand{\mtbf}{\ensuremath{\omega}}

\newcommand{\locationEnergy}[1]{\ensuremath{a_{#1}}}
\newcommand{\locationEnergyVector}{\ensuremath{\mathbf{\locationEnergy{}}}}
\newcommand{\budgetFunction}[1]{\ensuremath{B_k(#1)}}
\newcommand{\nodeHardwareExpenditure}[1]{\ensuremath{X_{#1}}}
\newcommand{\energyExpenditure}[1]{\ensuremath{Y_{#1}}}
\newcommand{\laborExpenditure}[1]{\ensuremath{Z_{#1}}}
\newcommand{\powerFunction}[1]{\ensuremath{P\left({#1}\right)}}
\newcommand{\informationFlowRate}[2]{\ensuremath{r_{#1#2}}}

\newcommand{\energyCost}{\ensuremath{\alpha}}
\newcommand{\exponentialConstant}{\ensuremath{C}}
\newcommand{\nodeCost}[1]{\ensuremath{\beta_{#1}}}
\newcommand{\failureRate}[1]{\ensuremath{\lambda_{#1}}}
\newcommand{\transmitPowerFunction}[2]{\ensuremath{T_{#1}(\distance{#1}{#2})}}
\newcommand{\optimalLifetime}[1]{\ensuremath{l_{#1}^{*}}}

\newcommand{\dataFlow}[2]{\ensuremath{b_{#1#2}}}
\newcommand{\isNodeAtLocation}[1]{\ensuremath{x_{#1}}}
\newcommand{\heaviside}[1]{\ensuremath{H[#1]}}
\newcommand{\maximumBatteryCapacity}{\ensuremath{D}}

\newcommand{\powerConsumption}{\ensuremath{\rho}}

\newcommand{\binaryExpenditure}[1]{\ensuremath{x_{#1}}}
\newcommand{\alternativePaymentVector}{\ensuremath{\mathbf{q}}}
\newcommand{\alternativeNPCFunction}[1]{\ensuremath{r(#1)}}
\newcommand{\maximumPayment}[1]{\ensuremath{\hat{\visitExpenditure{#1}}}}

\newcommand{\lifetimeFunctionYIntercept}[1]{\ensuremath{b_{#1}}}
\newcommand{\expenditureVector}{\ensuremath{\mathbf{\visitExpenditure{}}}}
\newcommand{\npcVariable}{\ensuremath{g}}
\newcommand{\npcFunction}{\ensuremath{\npcVariable(\expenditureVector)}}
\newcommand{\slackVariable}[1]{\ensuremath{u_{#1}}}
\newcommand{\slackVariableVector}{\mathbf{\slackVariable{}}}
\newcommand{\lagrangian}{\ensuremath{\mathcal{L}(\expenditureVector,\slackVariableVector)}}
\newcommand{\zeroVector}{\ensuremath{\mathbf{0}}}
\newcommand{\nextPaymentTermOne}[1]{\ensuremath{q(#1)}}
\newcommand{\linearLifetimeFunction}[1]{\ensuremath{\lifetimeFunctionSlope\visitExpenditure{#1}+\lifetimeFunctionYIntercept{#1}}}

\newcommand{\npcExpenditureFunction}{\ensuremath{h(\visitNumber)}}
\newcommand{\hessian}{\ensuremath{\mathbf{H}}}
\newcommand{\npcFunctionOpt}{\ensuremath{g(\mathbf{p}^*)}}

\newcommand{\choiceSet}{\ensuremath{\mathcal{H}}}
\newcommand{\numberOfRedundantNodes}{\ensuremath{G}}
\newcommand{\numberOfNodes}{\ensuremath{N}}

\newcommand{\npcValueFunction}[1]{\ensuremath{\npcVariable(#1)}}
\newcommand{\singleExpenditureBenchmark}{\ensuremath{s}}

\newcommand{\signalTransmitPower}[2]{\ensuremath{P^{\text{TX}}_{#1#2}}}
\newcommand{\signalReceivePower}{\ensuremath{P^{\text{RX}}}}
\newcommand{\antennaGain}[1]{\ensuremath{G_{#1}}}
\newcommand{\wavelength}{\ensuremath{\lambda}}
\newcommand{\pathlossExponent}{\ensuremath{\gamma}}
\newcommand{\laborCost}{\ensuremath{\zeta}}

\newcommand{\energyPaymentRate}{\ensuremath{\phi}}

\begin{document}
\title {Minimizing the Net Present Cost of Deploying and Operating Wireless Sensor Networks}

\author{Kevin~Dorling,~\IEEEmembership{Student Member,~IEEE,}
        Geoffrey~G.~Messier,~\IEEEmembership{Member,~IEEE,}
        Stefan~Valentin,~\IEEEmembership{Member,~IEEE,}
        and~Sebastian~Magierowski,~\IEEEmembership{Member,~IEEE}
\thanks{K. Dorling and G. G. Messier are with the Department of Electrical and Computer Engineering, University of Calgary, Alberta, Canada (e-mail: \{kudorlin,
gmessier\}@ucalgary.ca).}
\thanks{S. Valentin is with the Mathematical and Algorithmic Sciences Lab, FRC, Huawei Technologies, France (e-mail: stefan.valentin@huawei.com).}
\thanks{S. Magierowski is with the Department of Electrical Engineering and
Computer Science, York University, Toronto, Ontario, Canada (e-mail:
magiero@cse.yorku.ca).}}

\copyrightstatement

\maketitle

\begin{abstract}
Minimizing the cost of deploying and operating a Wireless Sensor Network (WSN) involves deciding how to partition a budget between competing expenses such as node hardware, energy, and labor. Most commercial network operators account for interest rates in their budgeting exercises, providing a financial incentive to defer some costs until a later time.  In this paper, we propose a net present cost (NPC) model for WSN capital and operating expenses that accounts for interest rates. Our model optimizes the number, size, and spacing between expenditures in order to minimize the NPC required for the network to achieve a desired operational lifetime. In general this optimization problem is non-convex, but if the spacing between expenditures is linearly proportional to the size of the expenditures, and the number of maintenance cycles is known in advance, the problem becomes convex and can be solved to global optimality. If non-deferrable recurring costs are low, then evenly spacing the expenditures can provide near-optimal results. With the provided models and methods, network operators can now derive a payment schedule to minimize NPC while accounting for various operational parameters. The numerical examples show substantial cost benefits under practical assumptions.
\end{abstract}


\IEEEpeerreviewmaketitle

\section{Introduction}

\IEEEPARstart{W}{ireless} Sensor Networks (WSNs) are groups of nodes that collaboratively collect information on an area of interest. Their ability to reduce costs and save human lives by autonomously monitoring remote and potentially hazardous regions has made them an active area of research, with applications in smart agriculture, environmental monitoring, detecting faults in systems and structures, disaster monitoring, and battlefield surveillance \cite{Akyildiz2002}. Nodes consist of sensors and transceivers to gather data on their immediate surroundings and forward this data over an ad-hoc network structure to predefined locations for further processing.  In order to minimize the cost of covering an area of interest, nodes are designed with inexpensive hardware, implement low-power protocols such as ZigBee \cite{iris}, and may use scheduling \cite{Alfieri2007} and energy-minimizing routing \cite{Chang2004}. By utilizing robots \cite{Suzuki2010} and unmanned aerial vehicles \cite{Corke2004} to replenish energy and replace damaged nodes, a network operator may reduce labor costs while extending WSN lifetime.

A network operator allocates a limited budget to numerous tasks related to building and maintaining a sensor network: hardware must be purchased and deployed, batteries may require periodic replacement or recharging, and damaged nodes may need to be replaced. Allocating additional money to one part of the budget reduces available funds for the other parts; for example, adding nodes to a network increases the portion of the budget dedicated to node hardware, but decreases the money available for energy and labor. As discussed in \sref{sec:related-work}, a large body of research focuses on minimizing the individual WSN costs, such as focusing solely on node hardware costs or node energy costs, but only a few papers study how to minimize the overall cost when multiple different expenses are combined together.

Before undertaking a project or submitting a bid, companies often estimate that project's initial investment, known as Capital Expenditures (CAPEX), and recurring expenditures, called Operational Expenditures (OPEX). For a WSN, CAPEX includes the costs of node hardware, the initial node energy supplies, and the labor required to initially deploy the network. OPEX includes the cost of replacement node hardware, replacement batteries, and the labor required to perform maintenance on the network.

To improve the cost of a WSN we propose a framework for minimizing its Net Present Cost (NPC). NPC is similar to net present value \cite{Dominick2007}, except all cash flows are considered outflows instead of being either inflows or outflows. NPC combines CAPEX and OPEX into a single cost by taking interest rates into account. Purchases made in the future cost less, in terms of the present currency value, as the operator earns interest by collecting revenue from the network it has built and by investing money elsewhere. Our framework could be used, for example, by a network operator that wants to take advantage of interest rates to reduce the cost of energy in the future. This requires spending more on labor in the future to visit the network and deliver this less expensive energy to the nodes. Our NPC minimization framework would, in this case, find the optimal balance between the money saved on energy and the cost of labor required to deliver it to the network.

By deferring costs to take advantage of interest rates, NPC minimization can significantly reduce the total cost of a WSN compared to paying for all costs up-front. Our framework produces a schedule of expenditures that minimizes NPC; this schedule can be used by a network operator when estimating the budget of a WSN. General rules-of-thumb can be applied in certain scenarios to produce near-optimal budgets. When non-deferrable recurring costs such as labor costs are low, the sensitivity of NPC minimization to the number of maintenance visits performed is also low, meaning that performing maintenance as often as possible produces a near-optimal NPC. In addition, we show that when non-deferrable recurring costs are low, evenly spacing the maintenance visits can provide a near-optimal NPC.

We define a \emph{visit} as a time point where the network operator visits the network to perform scheduled maintenance, such as restoring energy to nodes in the network.  Each visit has an expenditure associated with it, referred to as the \emph{visit expenditure}, while the time until the next visit is called the \emph{visit lifetime}. The visit made to initially deploy the network is a CAPEX expenditure, while visits made after deployment to restore node energy are OPEX expenditures. The network operator can adjust the number of visits, as well as each visit expenditure and visit lifetime in order to minimize the NPC. These parameters are interrelated, so adjustments are not always straightforward; for example, increasing visit expenditures may increase visit lifetimes and reduce the total number of visits required to achieve a desired operational lifetime. This action is only worthwhile if the reduced number of visits compensates for higher cost per visit.

We propose a two-layer optimization framework for determining the number of visits, the visit expenditures, and the visit lifetimes required to minimize the NPC. The first layer of the framework is a non-convex optimization problem that maximizes visit lifetime when given a visit expenditure that is known in advance. The visit lifetime depends on the visit expenditure: the more money spent on energy, the longer the visit lifetime. Maximizing visit lifetime minimizes future costs by taking full advantage of interest rates. This optimization problem is used to derive a \emph{lifetime function} for each visit. The lifetime function represents the relationship between visit expenditure and the maximum visit lifetime that can be achieved with that visit expenditure. Each visit may have its own unique lifetime function to account for changes in node hardware, energy, and labor costs between visits. 

The second layer of the framework optimizes visit expenditures and the number of visits to minimize the NPC. It uses the lifetime function from the first layer of the optimization framework to calculate the optimal visit lifetime for each visit expenditure. This ensures that the visit expenditures found by the second layer provide the optimal visit lifetimes. We show that the optimization problem in the second layer is non-linear and non-convex, making it difficult to find a globally optimum solution.

To reduce the complexity of this non-convex problem, we show that if the total number of visits is fixed to $\totalNumberOfVisits$, and the lifetime function for each visit is piecewise linear (equal to zero until a certain visit expenditure and increasing linearly with slope $\lifetimeFunctionSlope$ afterwards), then NPC minimization will be convex and the solution found will be globally optimal. The slope $\lifetimeFunctionSlope$ will be equal for all visits, but the point where the function transitions from zero to an increasing linear function may differ. Assuming that a maximum of $\maximumNumberOfVisits$ visits are allowed, we provide a $\bigO{\maximumNumberOfVisits^3}$ algorithm for finding the number of visits that minimize the NPC. The lifetime functions are piecewise linear under a network model that assumes optimum data flows between nodes, optimum energy consumption, and that 1-connectivity is adequate.

Points in time where the network operator performs unscheduled maintenance on the network, due to unexpected events such as hardware faults or environmental damage, are considered when calculating the NPC. As such events may occur at any given moment, we do not know when an unscheduled repair will occur, so we approximate the NPC of unscheduled payments by assuming that failures occur periodically, with the length of each period equal to the Mean Time Between Failures (MTBF) \cite{Jones2006} of the network nodes. The NPC of unscheduled payments can be reduced by improving the reliability of the network. Doing so increases the cost of node hardware, either by purchasing more robust hardware or by adding redundant nodes in stand-by to \emph{node locations}, the locations of active nodes in the network. To balance cost and robustness, we provide a method of minimizing node hardware costs and the approximate NPC of unscheduled repairs. The method assumes every node has identical hardware and every node location is given the same number of redundant nodes in stand-by, and finds the best type of node hardware and level of redundancy out of a set of possible choices.

A brief survey of the literature relevant to minimizing deployment and operational costs of WSNs is provided in \sref{sec:related-work}, while our network and cost models are given in \sref{sec:network-and-cost-models}. \sref{sec:visit-lifetime-framework} discusses how to maximize visit lifetime for a given visit expenditure by optimally dividing the visit expenditure between node hardware, energy, and labor costs. \sref{sec:visit-lifetime-framework} also demonstrates that the lifetime function, the relationship between visit expenditure and visit lifetime, is piecewise linear under our network and cost models. A general NPC minimization framework is proposed in \sref{sec:multiple-deployment-optimization}, and the linear lifetime function found in \sref{sec:visit-lifetime-framework} is exploited to show that equally spacing visits can be a near-optimal technique for minimizing the NPC under our network model. \sref{sec:numerical-results} provides numerical results for the NPC minimization problem under practical assumptions. It discusses the effect of network lifetime, the interest rate, the MTBF of the network nodes, and the costs of node hardware, energy, and labor on the NPC of the network.

\section{Related Work}
\label{sec:related-work}

When minimizing the overall WSN cost, NPC minimization accounts for the costs of node hardware, energy, and labor. The majority of papers available in the literature tend to focus on only a single type of cost. Both \cite{Younis2008} and \cite{Younis2014} are survey papers that contain techniques for minimizing the node count, and therefore node hardware costs, when connecting disjoint networks. The survey papers \cite{EnergyConservationSurvey} and \cite{EnergySurvey} discuss techniques for reducing network energy consumption and therefore energy costs. Labor costs can be reduced by scheduling node replacement and recharging activities \cite{Tong2011} and automating maintenance processes via robots \cite{Suzuki2010} and unmanned aerial vehicles \cite{Corke2004}. While the techniques presented in these papers can reduce network cost, they are not useful for operators seeking to allocate a budget across multiple costs.

There is a significant body of work on optimizing the deployment and operational costs of cellular networks \cite{timus2005,timus2008,werner2008,timus2009}. The fundamental difference with respect to this paper is that, once deployed, the infrastructure of a cellular network is static while the topology of a WSN varies in time. This variation results from failed sensor nodes due to depleted batteries and the dynamics of ad-hoc routing. Compared to cellular networks, WSNs require a substantially different model to compute the deployment of energy (i.e. batteries) and maintenance cycles.
 
To the best of our knowledge, the only papers in the literature accounting for multiple types of costs in a WSN are the works by Misra et al. \cite{Misra2010} and Dutta et al. \cite{Dutta2011}. Compared to these papers, the NPC minimization approach in this paper is unique in that it determines the visit expenditure, visit lifetime, and number of visits while taking advantage of interest rates. The other research that combines multiple types of network costs together either ignore savings from interest rates or fail to optimize the spacing between visits, both of which are critically important components of creating a network budget.

Misra et al. \cite{Misra2010} provide a method of balancing the cost of WSN maintenance with network performance loss. Performance refers to the probability of the WSN detecting an event; this probability, and therefore performance, decreases as the number of failed nodes increases. This work uses a probabilistic model to predict the long-term cost of a network, and provides an algorithm for determining the optimal trade-off between node replacement cost and performance loss. Their technique focuses solely on OPEX, ignoring the CAPEX required to deploy the network, and does not factor in savings due to interest rates, meaning that reductions in maintenance costs from returns on investments or revenues generated by the network are not considered when determining the node replacement policy. 

Dutta et al. \cite{Dutta2011} provide a strategy that considers interest rates when determining when to replace nodes in a network to minimize maintenance costs. This research determines the optimal year in which to replace nodes and assumes that maintenance costs are known in advance. It does not find the optimal size of each visit expenditure nor the optimal spacing between visits, limiting its applicability to minimizing a WSN's budget. By determining the amount to spend on each visit, the number of visits, and the spacing between them to minimize NPC, we allow network operators to not only reduce the total WSN cost, but to also estimate the budget and cash flows of a WSN.

\section{Network and Cost Models}
\label{sec:network-and-cost-models}

In this section we provide a network model and a cost model for the NPC minimization problem. \sref{ssec:network-and-cost-models} describes the types of locations present in the area of interest, their properties, and how they relate to one another to form a WSN. In \sref{ssec:cost-model} we present a model for NPC that takes into consideration the number of visits, the visit expenditures, and the visit lifetimes.

\subsection{Network Model}
\label{ssec:network-and-cost-models}

Every network has a set of sensor node locations $\sensorLocationSet{}$, a set of sink locations $\sinkLocationSet{}$, and a set of potential or candidate relay locations $\candidateRelayLocationSet{}$. The union of these sets is referred to as the universal set $\universalSet{} = \sensorLocationSet{} \cup \sinkLocationSet{} \cup \candidateRelayLocationSet{}$. Relay nodes forward data from sensor nodes to the sink nodes, possibly over multiple hops. Not every candidate relay location will have a relay node placed on it; depending on factors such as the transmit range of other nodes and the network topology, certain candidate relay locations may be chosen over others. Sensor nodes, in addition to performing the same forwarding duties as relay nodes, collect and forward data from their on-board sensors. Sink nodes collect the data gathered by every sensor node in the network and store or process it. Sensor node and sink node locations are assumed to always have sensor and sink nodes respectively placed on them. 

Sink, sensor, and relay nodes placed on their respective locations consume energy and handle data while performing their roles. Each sensor node $i$ in $\sensorLocationSet{}$ generates data at a rate of $\dataGenerationRate{i}$\,bits/s and consumes energy at a rate of $\sensingPower{i}$\,J/bit when sensing data. Each sensor or relay on location $i$ in $\candidateRelayLocationSet{} \cup \sensorLocationSet{}$ consumes $\transmitPower{i}$\,J/bit when transmitting to a location that is distance $\distance{}{}$ meters away, and consumes $\receivePower{i}$\,J/bit while receiving data. An edge exists from node $i$ to another node $j$ if the distance between the nodes $\distance{i}{j}$ is less than the maximum transmit distance in meters.

When determining how many nodes to place, where to place them, and how much energy to allocate to the nodes in a network, the operator has specific goals in mind; here we assume the goals are to achieve an operational lifetime of exactly $\operationalLifetime$ years and 1-connectivity at minimum cost. While other factors such as latency and quality of service are important in a number of situations, we assume real-time data acquisition is not critical and that the network will be lightly loaded, so that such factors are not a priority.

\subsection{Cost Model}
\label{ssec:cost-model}

\begin{figure}[]
  \centering
      \includegraphics[scale=0.8]{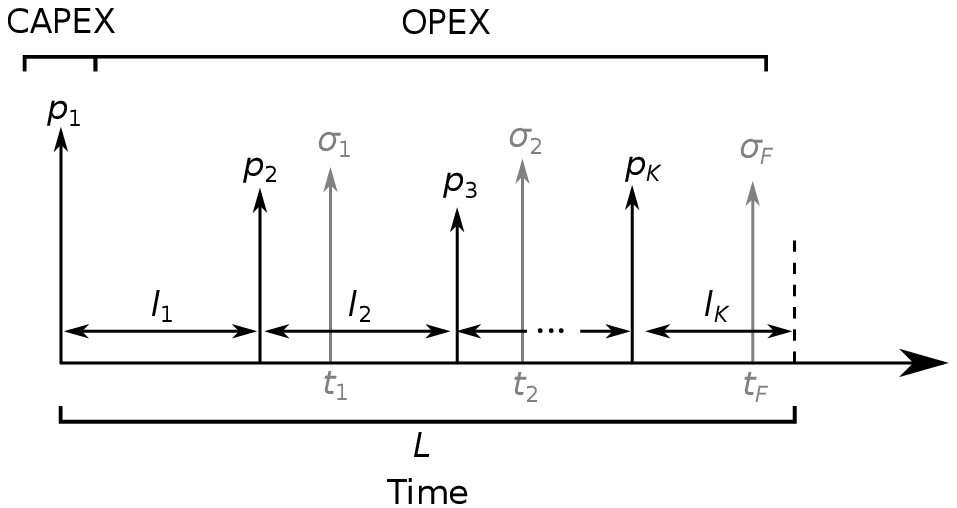}
  \caption{Payments made over the course of a network's lifetime. Scheduled payments $\visitExpenditure{1} \ldots \visitExpenditure{K}$ are known in advance, while unscheduled payments $\unscheduledPayment{1} \ldots \unscheduledPayment{\numberOfFailures}$ occur at times $\unscheduledPaymentTime{1} \ldots \unscheduledPaymentTime{\numberOfFailures}$.}
  \label{fig:npv-diagram}
\end{figure}

\fref{fig:npv-diagram} illustrates our cost model. We consider two types of payments: scheduled payments for performing routine tasks such as restoring energy to nodes, and unscheduled payments for reacting to unexpected issues such as hardware faults.

Scheduled payments are divided into events called visits. Assuming that a total number of $\totalNumberOfVisits$ visits are made, there is an initial visit expenditure $\visitExpenditure{1}$ that accounts for initially deploying the network, and a number of subsequent visit expenditures $\visitExpenditure{2} \ldots \visitExpenditure{\totalNumberOfVisits}$ for maintenance. The time between $\visitExpenditure{\visitNumber}$ and $\visitExpenditure{\visitNumber+1}$ is the visit lifetime $\visitLifetime{\visitNumber}$; as will be discussed in \sref{sec:visit-lifetime-framework}, the expenditure $\visitExpenditure{\visitNumber}$ is optimally divided between hardware, energy, and labor costs in such a way as to maximize $\visitLifetime{\visitNumber}$. The operational lifetime of the network is $\operationalLifetime = \sum_{1}^{\totalNumberOfVisits}\visitLifetime{\visitNumber}$. When minimizing the NPC of a network, one has to determine each visit expenditure $\visitExpenditure{\visitNumber}$, the total number of visits $\totalNumberOfVisits$, and the visit lifetimes $\visitLifetime{\visitNumber}$.

Unscheduled payments occur to handle network failures that cannot be prevented through routine maintenance, such as those caused by hardware faults, environmental hazards, or accidental damage. We assume that $\numberOfFailures$ failures occur at times $\unscheduledPaymentTime{1} \ldots \unscheduledPaymentTime{\numberOfFailures}$, with failures costing $\unscheduledPayment{1} \ldots \unscheduledPayment{\numberOfFailures}$ financial units each to repair.

Minimizing the cost of a network with a long operational lifetime entails determining whether money is best spent immediately, or later after it earns interest at a rate of $\interestRate>0$. In other words, one way to minimize costs is to minimize the Net Present Cost (NPC) of the network. Net present cost can be expressed as
\begin{align}
\label{eq:NPC}
\text{NPC} = 
\begin{cases}
\visitExpenditure{1} + \sum_{\unscheduledPaymentNumber=1}^{\numberOfFailures} \frac{\unscheduledPayment{\unscheduledPaymentNumber}}{(1+\interestRate)^{\unscheduledPaymentTime{\unscheduledPaymentNumber}}} & \totalNumberOfVisits = 1 \\
\visitExpenditure{1} + \sum_{\visitNumber = 2}^{\totalNumberOfVisits} \frac{\visitExpenditure{\visitNumber}}{(1+\interestRate)^{\sum_{\visitExpenditureIndex = 1}^{\visitNumber-1}\visitLifetime{\visitExpenditureIndex}}} + \sum_{\unscheduledPaymentNumber=1}^{\numberOfFailures} \frac{\unscheduledPayment{\unscheduledPaymentNumber}}{(1+\interestRate)^{\unscheduledPaymentTime{\unscheduledPaymentNumber}}} & \totalNumberOfVisits \ge 2.
\end{cases} 
\end{align}

We assume that a relationship exists between the visit expenditure $\visitExpenditure{\visitNumber}$ during visit $\visitNumber$, and the visit lifetime $\visitLifetime{\visitNumber}$. In other words
\begin{equation}
\visitLifetime{\visitNumber} = \visitLifetimeFunction{\visitNumber}{\visitExpenditure{\visitNumber}}.
\label{eq:lifetime}
\end{equation}
Each lifetime function $\visitLifetimeFunction{\visitNumber}{\visitExpenditure{\visitNumber}}$ may be unique for each visit $\visitNumber$.

As we do not know when failures occur, we propose a method to approximate failure times for the purpose of this analysis. We assume that the network's Mean Time Between Failures (MTBF) $\mtbf$ is known, that $\mtbf$ does not vary over the network's operational lifetime $\operationalLifetime$, and that the cost $\unscheduledPayment{}$ of failures does not change. As discussed at the end of \sref{ssec:network-and-cost-models}, we assume that 1-connectivity is adequate for the network. In this case, a single node failure potentially disconnects the network, meaning that $\mtbf$ is the same as the MTBF of the network nodes. With our approximation, $\unscheduledPaymentTime{\unscheduledPaymentNumber} = \unscheduledPaymentNumber \mtbf$ and $\unscheduledPayment{\unscheduledPaymentNumber} = \unscheduledPayment{}$, $\forall \unscheduledPaymentNumber \in [1,\lfloor \operationalLifetime/\mtbf \rfloor]$, meaning that the NPC can now be written as
\begin{equation}
\text{NPC} = 
\begin{cases}
\visitExpenditure{1} + \sum_{\unscheduledPaymentNumber=1}^{\lfloor \operationalLifetime / \mtbf \rfloor } \frac{ \unscheduledPayment{} }{ (1+\interestRate)^{ \unscheduledPaymentNumber \mtbf}} & \totalNumberOfVisits = 1 \\
\visitExpenditure{1} + \sum_{\visitNumber = 2}^{\totalNumberOfVisits} \frac{\visitExpenditure{\visitNumber}}{(1+\interestRate)^{\sum_{\visitExpenditureIndex = 0}^{\visitNumber-1}\visitLifetimeFunction{\visitExpenditureIndex}{\visitExpenditure{\visitExpenditureIndex}}}} & \\ \ \ \ \, + \sum_{\unscheduledPaymentNumber=1}^{\lfloor \operationalLifetime / \mtbf \rfloor} \frac{\unscheduledPayment{}}{(1+\interestRate)^{ \unscheduledPaymentNumber \mtbf}} & \totalNumberOfVisits \ge 2.
\end{cases}
\label{eq:NPC-with-lifetime}
\end{equation}

It is important to point out that minimizing overall NPC requires two layers of optimization.  In \sref{sec:visit-lifetime-framework}, a first optimization is used to determine how an expenditure $\visitExpenditure{\visitNumber}$ is optimally divided between hardware, energy, and labor costs in order to maximize $\visitLifetime{\visitNumber}$.  The results of this optimization establishes the function $\visitLifetime{\visitNumber} = \visitLifetimeFunction{\visitNumber}{\visitExpenditure{\visitNumber}}$. In \sref{sec:multiple-deployment-optimization} we then use the function $\visitLifetimeFunction{\visitNumber}{\visitExpenditure{\visitNumber}}$ in a second optimization that minimizes overall NPC by optimizing the visit expenditures $\visitExpenditure{\visitNumber}$.

\section{Framework for Optimizing Visit Lifetime}
\label{sec:visit-lifetime-framework}

From \eref{eq:NPC} we see that minimizing NPC involves maximizing visit lifetime, $\visitLifetime{\visitNumber}$.  We also assume in \eref{eq:lifetime} that a relationship $\visitLifetime{\visitNumber} = \visitLifetimeFunction{\visitNumber}{\visitExpenditure{\visitNumber}}$ exists.  In this section we provide an optimization problem that maximizes $\visitLifetime{\visitNumber}$ for a given $\visitExpenditure{\visitNumber}$ by optimizing the information flow rates in the network and the division of $\visitExpenditure{\visitNumber}$ between hardware, energy, and labor costs. This optimization is formulated as a non-linear, non-convex, yet continuous problem in \sref{ssec:optimization-problems}, and is turned into a non-continuous Mixed Integer Program (MIP) in \sref{ssec:mixed-integer-program-formulation} that is compatible with MIP solvers such as CPLEX \cite{cplex}.  Finally, it is shown in \sref{ssec:payment-lifetime-relationship} that the relationship $\visitLifetime{\visitNumber} = \visitLifetimeFunction{\visitNumber}{\visitExpenditure{\visitNumber}}$ is linear.

\subsection{The Visit Lifetime Maximization Problem}
\label{ssec:optimization-problems}

In this section we propose a continuous, non-linear optimization problem for maximizing the visit lifetime $\visitLifetime{\visitNumber}$ when given a visit expenditure $\visitExpenditure{\visitNumber}$, assuming the network model in \sref{ssec:network-and-cost-models}. By definition the visit number $\visitNumber \ge 1$. We let $\locationEnergyVector=\left[\locationEnergy{1} \  \locationEnergy{2} \  \ldots \ \locationEnergy{|\universalSet{}|} \right]$, where $\locationEnergy{i}$ is the energy allocated to each location $i \in \universalSet{}$. We express our budget $\budgetFunction{\locationEnergyVector}$ in terms of $\locationEnergyVector$. As we will show below, the energy vector can be used to determine the node hardware expenditure, $\nodeHardwareExpenditure{\visitNumber}$, and the energy expenditure, $\energyExpenditure{\visitNumber}$.  The sum of these costs with the labor expenditure, $\laborExpenditure{\visitNumber}$, is equal to the overall budget. By constraining the budget $\budgetFunction{\locationEnergyVector}$ to the visit expenditure $\visitExpenditure{\visitNumber}$, we can optimize the node hardware, energy, and labor expenditures without causing the budget to exceed $\visitExpenditure{\visitNumber}$. We also provide the power $\powerFunction{i}$ in Watts consumed at each location $i$ in terms of the information flow rate $\informationFlowRate{i}{j}$, the rate at which each location $i$ sends data to other locations $j$. After discussing the expressions for $\budgetFunction{\locationEnergyVector}$ and $\powerFunction{i}$, we propose the optimization problem itself.

The budget is a function of the location energy allocation vector $\locationEnergyVector$. The energy expenditure $\energyExpenditure{\visitNumber} = \sum_{i \in \universalSet{} }\energyCost \locationEnergy{i}$, which requires the energy $\locationEnergy{i}$ in Joules allocated to each location $i \in \universalSet{}$, as well as the cost $\energyCost$ of a single Joule. To determine whether or not a node is required at location $i$, we notice that location $i$ requires a node if $\locationEnergy{i} > 0$. We use an indicator function to tell us whether location $i$ needs a node based on the value of $\locationEnergy{i}$. To obtain a continuous formulation, the exponential function $\left(1-\exp(-\exponentialConstant \locationEnergy{i})\right)$ can be used as an indicator function. The constant $\exponentialConstant$ is a large number such that the function is approximately 1 when $\locationEnergy{i} > 0$, and 0 otherwise. 

We assume that all nodes have identical hardware, and are purchased at a price of $\nodeCost{}$ financial units per node during the initial visit. We therefore let $\nodeHardwareExpenditure{\visitNumber} = \sum_{i \in \universalSet{}} \nodeCost{} \left( 1 - \exp(-\exponentialConstant \locationEnergy{i}) \right)$ when $k = 1$, and $\nodeHardwareExpenditure{\visitNumber} = 0$ when $\visitNumber \ge 2$. In this case, all nodes are assumed to have the same cost $\nodeCost{}$ and the same rate of failure $\failureRate{}$, regardless of whether they are sensor, relay, or sink nodes.

As we assume that node hardware is identical and that 1-connectivity is sufficient, each node has the same failure rate $\failureRate{}$, and a single node failure potentially results in network failure. The overall Mean Time Between Failures (MTBF) of the network is therefore
\begin{equation}
\mtbf = \frac{1}{\sum_{i \in \universalSet{}} \failureRate{}\left(1-\exp(-\exponentialConstant \locationEnergy{i})\right)} = \frac{\nodeCost{}}{\failureRate{} \nodeHardwareExpenditure{1}}.
\end{equation}
By optimizing the node hardware expenditure $\nodeHardwareExpenditure{1}$ during the first visit, we are also maximizing the MTBF $\mtbf$. We do not consider placing redundant nodes in this section, as that is covered in the method for optimizing node hardware costs and the approximate NPC of unscheduled payments in \sref{ssec:minimizing-unscheduled-npc}.

We assume that the labor expenditure $\laborExpenditure{\visitNumber}$ during visit $\visitNumber$ is known in advance. Later in this section we show that, because we optimally allocate energy to the nodes, every location with a node will be visited when maintenance is performed. This means we know which locations to visit in advance, and can estimate the time required to travel between the different locations, and therefore the time spent and cost of performing labor. When factoring in node hardware, energy, and labor costs, our budget can be expressed as
\begin{equation}
\label{eq:budget}
\budgetFunction{\locationEnergyVector} = 
\begin{cases}
\sum_{i \in \universalSet{} } \left( \nodeCost{} \left(1-\exp(-\exponentialConstant \locationEnergy{i})\right) + \energyCost \locationEnergy{i}\right) + \laborExpenditure{\visitNumber} & \visitNumber = 1 \\
\sum_{i \in \universalSet{}} \left(\energyCost \locationEnergy{i}\right) + \laborExpenditure{\visitNumber} & \visitNumber \ge 2.
\end{cases}
\end{equation}

The power $\powerFunction{i}$ in Watts consumed by node $i \in \universalSet{}$ can be expressed in terms of the information flow rates $\informationFlowRate{i}{j}$. Location $i$ sends data to location $j$ at a rate of $\informationFlowRate{i}{j}$ bits/s. We let the sets $\candidateRelayLocationSet{i}$, $\sensorLocationSet{i}$, and $\universalSet{i}$ represent the set of nodes in $\candidateRelayLocationSet{}$, $\sensorLocationSet{}$, and $\universalSet{}$ respectively that are within communication range of node $i$. The energy consumed by node $i$ when transmitting to another node $j$ is $\transmitPowerFunction{i}{j}$\,J/bit, where $\distance{i}{j}$ refers to the distance between the nodes in m. Node $i$ consumes $\receivePower{i}$\,J/bit while receiving data. Sensor $i$ consumes $\sensingPower{i}$\,J/bit collecting data at a rate of $\dataGenerationRate{i}$\,bits/s. The expression for the power in Watts consumed by a node is
\begin{align}
\powerFunction{i} = & \sum_{j \in \universalSet{i}} \transmitPowerFunction{i}{j} \informationFlowRate{i}{j} + \sum_{j \in \candidateRelayLocationSet{i} \cup \sensorLocationSet{i}} \receivePower{i} \informationFlowRate{j}{i} + \sensingPower{i} \dataGenerationRate{i}.
\label{eq:energy-function}
\end{align}

We can use \eref{eq:budget} and \eref{eq:energy-function} to formulate the lifetime maximization problem as a non-linear, non-convex, yet continuous optimization problem
\begin{subequations}
\label{eq:lifetime-optimization}
\begin{align}
\max \limits_{\locationEnergy{i},\informationFlowRate{i}{j},\visitLifetime{\visitNumber} \, \in \, \mathbb{R}^+} & \visitLifetime{\visitNumber} \label{eq:lifetime-maximization-objective} \\
\underset{\hphantom{\max \limits_{\locationEnergy{i},\informationFlowRate{i}{j},\visitLifetime{\visitNumber} \, \in \, \mathbb{R}^+}}}{\text{s.t.}} & \sum_{j \in \candidateRelayLocationSet{i} \cup \sensorLocationSet{i}} \informationFlowRate{j}{i} + \dataGenerationRate{i} = \sum_{j \in \universalSet{i}} \informationFlowRate{i}{j} & \forall i \in \candidateRelayLocationSet{} \cup \sensorLocationSet{} \label{eq:lifetime-maximization-flow-constraint} \\
& \visitLifetime{\visitNumber} \powerFunction{i} = \locationEnergy{i} \label{eq:lifetime-maximization-energy-constraint} & \forall i \in \universalSet{} \\
&\budgetFunction{\locationEnergyVector} = \visitExpenditure{\visitNumber}. \label{eq:lifetime-maximization-budget}
\end{align}
\end{subequations}
The energy $\locationEnergy{i}$ allocated to node $i$, the information flow rate $\informationFlowRate{i}{j}$ between node $i$ and node $j$, and the visit lifetime $\visitLifetime{\visitNumber}$ are all optimization variables that exist in the non-negative reals. The elements of vector $\locationEnergyVector$ are the energy allocation values $\locationEnergy{i}$. Flow constraint \eref{eq:lifetime-maximization-flow-constraint} ensures each node does not transmit more data per second than it receives from others or collects through sensing. The sensor data generation rate $\dataGenerationRate{i} > 0, \forall i \in \sensorLocationSet{}$\, because sensors generate data, while $\dataGenerationRate{i} = 0, \forall i \in \candidateRelayLocationSet{}$\, because relays do not. Energy constraint \eref{eq:lifetime-maximization-energy-constraint} ensures each node $i$ uses all of the $\locationEnergy{i}$ Joules allocated to it; we explain later in this section why this is an equality constraint. Budget constraint \eref{eq:lifetime-maximization-budget} lets us find the optimal node hardware, energy, and labor expenditures resulting from adjustments to $\locationEnergyVector$, while ensuring that the budget does not exceed the visit expenditure $\visitExpenditure{\visitNumber}$. The left-hand sides of \eref{eq:lifetime-maximization-energy-constraint} and \eref{eq:lifetime-maximization-budget} are non-linear; as both constraints are equality constraints, \eref{eq:lifetime-optimization} is non-convex.

The energy constraint \eref{eq:lifetime-maximization-energy-constraint} is typically defined as an inequality constraint (for example, in \cite{Alfieri2007,Chang2004}, and \cite{Cheng2008}). To demonstrate why representing  \eref{eq:lifetime-maximization-energy-constraint} as an equality constraint is valid when maximizing lifetime and optimally allocating energy, we will prove by contradiction that at optimality the left-hand side of \eref{eq:lifetime-maximization-energy-constraint} must be equivalent to the right-hand side. Suppose that at the optimal lifetime $\optimalLifetime{\visitNumber}$, $\optimalLifetime{\visitNumber}\powerFunction{i} < \locationEnergy{i}$ for a node $i$. This implies that
\begin{enumerate}
\item At optimal lifetime $\optimalLifetime{\visitNumber}$ one or more nodes have exhausted their supply of energy.
\item Node $i$ has $\left(\locationEnergy{i} - \optimalLifetime{\visitNumber}\powerFunction{i}\right)$\,J of energy that it has not yet spent.
\end{enumerate}
If, when adding or replacing batteries, a portion of the spare energy in node $i$ had instead been allocated to the nodes with no energy, a lifetime longer than $\optimalLifetime{\visitNumber}$ would have been achieved. This means that $\optimalLifetime{\visitNumber}$ is not in fact optimal if $\exists i \in \universalSet{}: \optimalLifetime{\visitNumber}\powerFunction{i} < \locationEnergy{i}$. Therefore $\optimalLifetime{\visitNumber}$ is only optimal as long as $\optimalLifetime{\visitNumber}\powerFunction{i} = \locationEnergy{i}, \forall i \in \universalSet{}$. This further implies that all locations with nodes will be visited when replacing batteries, as all nodes will have exhausted their energy supplies at the optimal lifetime $\optimalLifetime{\visitNumber}$.

The same argument can be used to justify making the budget constraint \eref{eq:lifetime-maximization-budget} an equality constraint. We can prove by contradiction that, at the optimal lifetime $\optimalLifetime{\visitNumber}$, the budget $\budgetFunction{\locationEnergyVector} = \visitExpenditure{\visitNumber}$. Suppose that at the optimal lifetime $\optimalLifetime{\visitNumber}$, $\budgetFunction{\locationEnergyVector} < \visitExpenditure{\visitNumber}$, meaning that we have extra money $\left(\visitExpenditure{\visitNumber} - \budgetFunction{\locationEnergyVector}\right)$ that has not been spent. Additional energy could have been purchased with this money, increasing $\locationEnergy{i}, \forall i \in \universalSet{}$ until $\budgetFunction{\locationEnergyVector} = \visitExpenditure{\visitNumber}$. From \eref{eq:lifetime-maximization-flow-constraint} we know that the flow rates $\informationFlowRate{i}{j}$ will not increase with energy $\locationEnergy{i}$: the rate $\dataGenerationRate{i}$ at which data is generated by sensor $i$ is fixed, therefore according to the energy constraint \eref{eq:lifetime-maximization-energy-constraint} lifetime $\optimalLifetime{\visitNumber}$ must increase. This implies that $\optimalLifetime{\visitNumber}$ is not in fact optimal, meaning an optimal lifetime $\optimalLifetime{\visitNumber}$ requires that $\budgetFunction{\locationEnergyVector} = \visitExpenditure{\visitNumber}$.

\subsection{Formulation of the Mixed Integer Program}
\label{ssec:mixed-integer-program-formulation}

The optimization problem given in \eref{eq:lifetime-optimization} is non-convex, making it difficult for a solver to find a globally optimal solution. Even though \eref{eq:lifetime-optimization} is non-convex, it can be re-written as a Mixed Integer Program (MIP) and can therefore be solved with the robust branch-and-bound algorithms and heuristics available in commercial MIP solvers. The constraints will all be made linear, making the problem compatible with MIP solvers such as CPLEX \cite{cplex}.

Multiplying \eref{eq:lifetime-maximization-flow-constraint} by $\visitLifetime{\visitNumber}$ allows us to express the flow of data between node $i$ and node $j$ as $\dataFlow{i}{j}$\,bits instead of the rate $\informationFlowRate{i}{j}$\,bits/s. The left-hand side of \eref{eq:lifetime-maximization-energy-constraint} is re-written in terms of $\dataFlow{i}{j}$. To turn the continuous formulation \eref{eq:lifetime-optimization} into a MIP, we replace the indicator function with the binary optimization variable $\isNodeAtLocation{i}$ that is 1 when the node either exists at or will be added to candidate relay location $i$, and 0 otherwise. With these changes in mind, the optimization problem \eref{eq:lifetime-optimization} can be rewritten as
\begin{subequations}
\label{eq:mip-lifetime-optimization}
\begin{align}
\max \limits_{\substack{\locationEnergy{i},\dataFlow{i}{j},\visitLifetime{\visitNumber} \, \in \, \mathbb{R}^+\\ \isNodeAtLocation{i} \in \{0,1\}}} & \visitLifetime{\visitNumber} \label{eq:mip-lifetime-maximization-objective} \\
\underset{\hphantom{\max \limits_{\locationEnergy{i},\dataFlow{i}{j},\visitLifetime{\visitNumber} \, \in \, \mathbb{R}^+}}}{\text{s.t.}} & \sum_{j \in \candidateRelayLocationSet{i} \cup \sensorLocationSet{i}} \dataFlow{j}{i} + \dataGenerationRate{i} \visitLifetime{\visitNumber} = \sum_{j \in \universalSet{i}} \dataFlow{i}{j} & \forall i \in \candidateRelayLocationSet{} \cup \sensorLocationSet{} \label{eq:mip-lifetime-maximization-flow-constraint} \\
& \sum_{j \in \universalSet{i}} \transmitPowerFunction{i}{j} \dataFlow{i}{j} + \sum_{j \in \candidateRelayLocationSet{i} \cup \sensorLocationSet{i}} \receivePower{i} \dataFlow{j}{i} \nonumber \\
& \ \ \ \ + \sensingPower{i} \dataGenerationRate{i} \visitLifetime{\visitNumber} = \locationEnergy{i} \label{eq:mip-lifetime-maximization-energy-constraint} & \forall i \in \universalSet{} \\
& \sum_{i \in \universalSet{} } \left( \nodeCost{} \isNodeAtLocation{i} \left(1-\heaviside{\visitNumber-2}\right) + \energyCost \locationEnergy{i}\right) \nonumber \\
& \ \ \ \ + \laborExpenditure{\visitNumber} = \visitExpenditure{\visitNumber} \label{eq:mip-lifetime-maximization-budget} \\
& \locationEnergy{i} \le \maximumBatteryCapacity \isNodeAtLocation{i} & \forall i \in \universalSet{} \label{eq:mip-lifetime-maximization-energy-limit} \\
& \isNodeAtLocation{i} = 1 & \forall i \in \sensorLocationSet{} \cup \sinkLocationSet{}, \label{eq:mip-lifetime-maximization-node-exists} 
\end{align}
\end{subequations}
where $\maximumBatteryCapacity$ is a constant greater than or equal to the largest possible battery capacity. The flow, energy, and budget constraints \eref{eq:mip-lifetime-maximization-flow-constraint}, \eref{eq:mip-lifetime-maximization-energy-constraint}, and \eref{eq:mip-lifetime-maximization-budget} serve the same purposes as their counterparts \eref{eq:lifetime-maximization-flow-constraint}, \eref{eq:lifetime-maximization-energy-constraint}, and \eref{eq:lifetime-maximization-budget} respectively. Note that $\heaviside{\visitNumber-2}$ represents the Heaviside step function, making $\nodeCost{} \isNodeAtLocation{i} \left(1-\heaviside{\visitNumber-2}\right)=\nodeCost{} \isNodeAtLocation{i}$ when $\visitNumber=1 $, and $0$ otherwise. This is done because, as discussed in \sref{ssec:optimization-problems}, nodes are only purchased during the initial visit. We ensure that energy is only added to location $i$ if it has a node via \eref{eq:mip-lifetime-maximization-energy-limit}. Constraint \eref{eq:mip-lifetime-maximization-node-exists} ensures that sensor and sink nodes are always placed on the field. The objective and all the constraints in \eref{eq:mip-lifetime-optimization} are in a linear form; note that the Heaviside function is solved prior to optimization, meaning that \eref{eq:mip-lifetime-maximization-budget} is linear. The optimization variables are either real numbers or integers, making \eref{eq:mip-lifetime-optimization} compatible with commercial MIP optimizers.

\subsection{Lifetime Function Derivation}
\label{ssec:payment-lifetime-relationship}

To minimize the overall NPC in \sref{sec:multiple-deployment-optimization}, it is necessary to express the results of the $\visitLifetime{\visitNumber}$ maximization in this section in terms of the $\visitLifetime{\visitNumber} = \visitLifetimeFunction{\visitNumber}{\visitExpenditure{\visitNumber}}$ function in (2). By doing so, we demonstrate that $\visitLifetimeFunction{\visitNumber}{\visitExpenditure{\visitNumber}}$ is a linear function.

After running the optimization described in the previous sections, the maximum visit lifetime $\visitLifetime{\visitNumber}^*$, and the optimal energy $\locationEnergy{i}^*$ at each node $i$ that achieves it, have been found. The budget function \eref{eq:budget} can therefore be rewritten as
\begin{equation}
\label{eq:budget2}
\budgetFunction{\locationEnergyVector^*} = \nodeHardwareExpenditure{\visitNumber} + \energyCost  \visitLifetime{\visitNumber}^* \powerConsumption + \laborExpenditure{\visitNumber},
\end{equation}
where $\locationEnergyVector^*$ is a vector whose element $i$ is $\locationEnergy{i}^*$, and $\powerConsumption = \sum_{i \in \universalSet{} } \powerFunction{i}$ represents the overall network power consumption after the optimal information flow rates $\informationFlowRate{i}{j}^*$ have been determined. Constraint \eref{eq:lifetime-maximization-energy-constraint} allows us to substitute $\locationEnergy{i}^*$ with $\visitLifetime{\visitNumber}^* \powerFunction{i}$, therefore $\energyCost \visitLifetime{\visitNumber}^* \powerConsumption = \energyCost  \sum_{i \in \universalSet{} } \left(\visitLifetime{\visitNumber}^* \powerFunction{i}\right) = \energyCost \sum_{i \in \universalSet{} } \locationEnergy{i}^*$. The node hardware expenditure $\nodeHardwareExpenditure{\visitNumber} = \nodeCost{} \sum_{i \in \universalSet{} } \left(1-\exp(-\exponentialConstant \locationEnergy{i}^*)\right)$ when $\visitNumber=1$, and $\nodeHardwareExpenditure{\visitNumber} = 0$ otherwise.

Rearranging \eref{eq:budget2} and letting $\budgetFunction{\locationEnergyVector^*} = \visitExpenditure{\visitNumber}$ as in \eref{eq:lifetime-maximization-budget} gives
\begin{equation}
\visitLifetime{\visitNumber}^* = \visitLifetimeFunction{\visitNumber}{\visitExpenditure{\visitNumber}} =
\begin{cases}
\frac{\visitExpenditure{\visitNumber} - \nodeHardwareExpenditure{\visitNumber} - \laborExpenditure{\visitNumber}  }{\energyCost \powerConsumption} & \visitExpenditure{\visitNumber} > \nodeHardwareExpenditure{\visitNumber} + \laborExpenditure{\visitNumber} \\
0 & \visitExpenditure{\visitNumber} \le \nodeHardwareExpenditure{\visitNumber} + \laborExpenditure{\visitNumber}.
\end{cases}
\label{eq:linear-relationship}
\end{equation}
The terms $\nodeHardwareExpenditure{\visitNumber}$, $\laborExpenditure{\visitNumber}$,\,$\energyCost$, and $\powerConsumption$ do not change with $\visitExpenditure{\visitNumber}$. From \eref{eq:lifetime-maximization-flow-constraint}, the flow rates $\informationFlowRate{i}{j}$ depend on the sensor data generation rate $\dataGenerationRate{i}$, which is constant. From \eref{eq:energy-function} we can see that $\powerFunction{i}$ and therefore $\powerConsumption$ only change with $\informationFlowRate{i}{j}$ and $\dataGenerationRate{i}$, so $\powerConsumption$ is not a function of $\visitExpenditure{\visitNumber}$. The flow rates and therefore the locations requiring nodes do not change with $\visitExpenditure{\visitNumber}$, so $\nodeHardwareExpenditure{\visitNumber}$ does not change with $\visitExpenditure{\visitNumber}$. Both $\laborExpenditure{\visitNumber}$ and $\energyCost$ are constants. As a result, $\visitLifetime{\visitNumber}^*$ increases linearly with $\visitExpenditure{\visitNumber}$.

\section{Framework for Optimizing Net Present Cost}
\label{sec:multiple-deployment-optimization}

When $\visitLifetime{\visitNumber}$ is maximized for a given $\visitExpenditure{\visitNumber}$, we established in \sref{sec:visit-lifetime-framework} that the relationship $\visitLifetime{\visitNumber} = \visitLifetimeFunction{\visitNumber}{\visitExpenditure{\visitNumber}}$ is linear. In this section we utilize this relationship to minimize the overall WSN NPC expressed in \eref{eq:NPC-with-lifetime}.

\sref{ssec:general-npc-minimization-framework} provides a general non-linear, non-convex NPC minimization formulation that is applicable to any $\visitLifetime{\visitNumber} = \visitLifetimeFunction{\visitNumber}{\visitExpenditure{\visitNumber}}$ lifetime function, linear or not. This generalized formulation is non-convex and difficult to solve optimally. However, in \sref{ssec:npc-minimization-with-linear-relationship} we demonstrate that when the linear lifetime function $\visitLifetime{\visitNumber} = \visitLifetimeFunction{\visitNumber}{\visitExpenditure{\visitNumber}}$ defined in \sref{sec:visit-lifetime-framework} is used, NPC minimization for a fixed number of visits $\totalNumberOfVisits$ is convex. Assuming that a maximum of $\maximumNumberOfVisits$ visits may be made, we provide a $\mathcal{O}(\maximumNumberOfVisits^3)$ algorithm for minimizing the NPC with a linear lifetime function. In \sref{ssec:equal-visit-lifetime-approximation} we show that equally spacing the visits can be a good rule of thumb when $\visitLifetimeFunction{\visitNumber}{\visitExpenditure{\visitNumber}}$ is linear.

The NPC minimization framework assumes that the network's Mean Time Between Failures (MTBF) is known and constant. The network's MTBF and therefore the NPC of unscheduled payments might depend, however, on the cost of node hardware: using more expensive nodes or adding redundant nodes in stand-by to each node location may improve the MTBF. A method of balancing the initial cost of node hardware with the approximate NPC of unscheduled payments is provided in \sref{ssec:minimizing-unscheduled-npc}.

\subsection{General NPC Minimization Framework}
\label{ssec:general-npc-minimization-framework}

Our goal is to minimize the NPC in \eref{eq:NPC-with-lifetime} while ensuring that the network remains operational for exactly $\operationalLifetime$ years. As discussed in \sref{ssec:cost-model}, the relationship between visit expenditure $\visitExpenditure{\visitNumber}$ and visit lifetime $\visitLifetime{\visitNumber}$ of visit $\visitNumber$ is the function $\visitLifetimeFunction{\visitNumber}{\visitExpenditure{\visitNumber}}$. By finding the optimum visit expenditures $\visitExpenditure{\visitNumber}^*$ we find the optimum visit lifetimes $\visitLifetime{\visitNumber}^* = \visitLifetimeFunction{\visitNumber}{\visitExpenditure{\visitNumber}^*}$ and thus the optimum spacing between visits.

To minimize the NPC, we require upper bounds on the visit expenditures and the number of visits. The maximum number of visits is $\maximumNumberOfVisits$, meaning $\totalNumberOfVisits \le \maximumNumberOfVisits$. For example, if the network operator could not feasibly visit the network more than once per month, then $\maximumNumberOfVisits=12\operationalLifetime$. The maximum visit expenditure for visit $\visitNumber$ is denoted $\hat{\visitExpenditure{\visitNumber}}$, meaning $\visitExpenditure{\visitNumber} \le \hat{\visitExpenditure{\visitNumber}}, \forall \visitNumber \in [1,\maximumNumberOfVisits]$. 

The NPC minimization problem for when $\maximumNumberOfVisits \ge 2$ can be written as
\begin{subequations}
\label{eq:npc-minimization}
\begin{align}
\min \limits_{\substack{\visitExpenditure{1} \ldots \visitExpenditure{\maximumNumberOfVisits} \, \in \, \mathbb{R}^+\\ \binaryExpenditure{\visitNumber},\binaryExpenditure{\visitNumber+1} \in \{0,1\}}} & \visitExpenditure{1} + \sum_{\visitNumber = 2}^{\maximumNumberOfVisits} \frac{\visitExpenditure{\visitNumber}}{\left(1+\interestRate\right)^{\sum_{\visitExpenditureIndex = 1}^{\visitNumber-1}{\visitLifetimeFunction{\visitExpenditureIndex}{\visitExpenditure{\visitExpenditureIndex}}}}} \nonumber \\ 
& \ \ \  + \sum_{\unscheduledPaymentNumber=1}^{\lfloor \operationalLifetime / \mtbf \rfloor} \frac{\unscheduledPayment{}}{(1+\interestRate)^{ \unscheduledPaymentNumber \mtbf}} & \label{eq:npc-minimization-objective} \\
\underset{\hphantom{\visitExpenditure{\binaryExpenditure{\visitNumber},\binaryExpenditure{\visitNumber+1} \in \{0,1\}}}}{s.t.} & \sum_{\visitNumber=1}^{\maximumNumberOfVisits} \visitLifetimeFunction{\visitNumber}{\visitExpenditure{\visitNumber}} = \operationalLifetime \label{eq:npc-minimization-lifetime} \\
& \visitExpenditure{\visitNumber} \le \binaryExpenditure{\visitNumber} \maximumPayment{\visitNumber} \label{eq:npc-minimization-payment} & \forall \visitNumber \in [1,\maximumNumberOfVisits]\\
& \binaryExpenditure{\visitNumber+1} \le \binaryExpenditure{\visitNumber} & \forall \visitNumber \in [1,\maximumNumberOfVisits-1] \label{eq:npc-minimization-payment-order} \\
& \visitLifetimeFunction{\visitNumber}{\visitExpenditure{\visitNumber}} \ge 0 & \forall \visitNumber \in [1,\maximumNumberOfVisits], \label{eq:npc-minimization-payment-lower-bound}
\end{align}
\end{subequations}
where the binary value $\binaryExpenditure{\visitNumber}$ represents whether or not a network operator performs visit $\visitNumber$, and  $\interestRate > 0$ represents the rate at which money earns interest. The constants $\mtbf$ and $\unscheduledPayment{}$ represent the network's MTBF and the cost of network failure respectively. Note that when $\maximumNumberOfVisits = 1$, finding the NPC is trivial: the visit expenditure $\visitExpenditure{1}$ that achieves the operational lifetime $\operationalLifetime$ is chosen. The objective function \eref{eq:npc-minimization-objective} is the NPC, while the lifetime constraint \eref{eq:npc-minimization-lifetime} ensures that the visit expenditures provide the required operational lifetime of $\operationalLifetime$. Constraint \eref{eq:npc-minimization-payment} makes $\visitExpenditure{\visitNumber} = 0$ when visit $\visitNumber$ is not made, while guaranteeing that $\visitExpenditure{\visitNumber}$ does not exceed the maximum expenditure amount $\maximumPayment{\visitNumber}$ when a visit is made. Constraint \eref{eq:npc-minimization-payment-order} is used to find the optimum number of visits: if $\binaryExpenditure{i} = 0$ for visit $i$, then $\visitExpenditure{j}=\binaryExpenditure{j}=0, \forall j \in [i,\maximumNumberOfVisits]$, making the optimum number of visits $\totalNumberOfVisits^* = \sum_{\visitNumber=1}^{\maximumNumberOfVisits} \binaryExpenditure{\visitNumber}$. Constraint \eref{eq:npc-minimization-payment-lower-bound} ensures that visit lifetime can never be negative.

The NPC minimization problem \eref{eq:npc-minimization} is a non-linear mixed integer program. If $\visitLifetimeFunction{\visitNumber}{\visitExpenditure{\visitNumber}}$ is non-linear, then \eref{eq:npc-minimization} is non-convex. Suppose that $\maximumNumberOfVisits \ge 2$, $\alternativePaymentVector = [ \visitExpenditure{1} \ \visitExpenditure{2} \ \ldots \ \visitExpenditure{\maximumNumberOfVisits}]$, and $\alternativeNPCFunction{\alternativePaymentVector}$ is \eref{eq:npc-minimization-objective}. It can be shown that
$$
\frac{\partial^2 \alternativeNPCFunction{\alternativePaymentVector}}{\partial \visitExpenditure{i} \partial \visitExpenditure{\maximumNumberOfVisits}} = \frac{\partial^2 \alternativeNPCFunction{\alternativePaymentVector}}{\partial \visitExpenditure{\maximumNumberOfVisits} \partial \visitExpenditure{i}} =
\begin{cases}
-\frac{\ln(1+\interestRate)}{(1+\interestRate)^{\sum_{\visitNumber = 1}^{\maximumNumberOfVisits-1}{\visitLifetimeFunction{\visitNumber}{\visitExpenditure{\visitNumber}}}}} \frac{\partial \visitLifetimeFunction{i}{\visitExpenditure{i}}}{\partial \visitExpenditure{i}} & i \neq \maximumNumberOfVisits \\
0 & i = \maximumNumberOfVisits .
\end{cases}
$$
We can see that the element at row  $\maximumNumberOfVisits$ and column $\maximumNumberOfVisits$ of the Hessian of $\alternativeNPCFunction{\alternativePaymentVector}$ is zero, and the other elements in row $\maximumNumberOfVisits$ and in column $\maximumNumberOfVisits$ are non-zero. The Hessian is therefore not positive semi-definite, meaning that the objective function \eref{eq:npc-minimization-objective} is non-convex. When $\maximumNumberOfVisits = 1$, finding the NPC is trivial because $\visitExpenditure{1}$ is set to the value that ensures $\visitLifetimeFunction{1}{\visitExpenditure{1}}$ equals operational lifetime $\operationalLifetime$.

\subsection{NPC Minimization with a Linear Lifetime Function}
\label{ssec:npc-minimization-with-linear-relationship}

Based on our proof of linearity of the lifetime function $\visitLifetimeFunction{\visitNumber}{\visitExpenditure{\visitNumber}}$ in \sref{sec:visit-lifetime-framework}, this section will demonstrate that the NPC minimization problem is convex for a fixed number of visits $\totalNumberOfVisits$. We provide an algorithm to determine the optimal number of visits in $\bigO{\maximumNumberOfVisits^3}$ time, for the maximal number of visits $\maximumNumberOfVisits$.

As shown in \sref{ssec:payment-lifetime-relationship}, we can assume the lifetime function has the form
$$
\visitLifetimeFunction{\visitNumber}{\visitExpenditure{\visitNumber}} =
\begin{cases}
\linearLifetimeFunction{\visitNumber}
& \visitExpenditure{\visitNumber} > -\lifetimeFunctionYIntercept{\visitNumber}/\lifetimeFunctionSlope \\
0 & \visitExpenditure{\visitNumber} \le -\lifetimeFunctionYIntercept{\visitNumber}/\lifetimeFunctionSlope,
\end{cases}
$$
where $\lifetimeFunctionSlope > 0$. Note that the slope $\lifetimeFunctionSlope$ is not a function of $\visitNumber$. A constant slope $\lifetimeFunctionSlope$ for all visits implies that the cost of energy, and the rate that the network consumes energy, do not change over time. Using \eref{eq:linear-relationship} from \sref{ssec:payment-lifetime-relationship}, the slope $\lifetimeFunctionSlope = \frac{1}{\energyCost \powerConsumption}$ and y-intercept $\lifetimeFunctionYIntercept{\visitNumber} = -\frac{\nodeHardwareExpenditure{\visitNumber}+\laborExpenditure{\visitNumber} }{\energyCost \powerConsumption}$. As with the general NPC minimization problem, we find the optimum visit lifetimes $\visitLifetime{\visitNumber}^* = \visitLifetimeFunction{\visitNumber}{\visitExpenditure{\visitNumber}^*}$ by first finding the optimum visit expenditures $\visitExpenditure{\visitNumber}^*$.

In this section we first provide a method for finding the visit expenditures $\visitExpenditure{\visitNumber}$ for $\visitNumber \in [1,\totalNumberOfVisits]$ to minimize NPC when $\totalNumberOfVisits$ is fixed. We then provide a $\bigO{\maximumNumberOfVisits^3}$ algorithm for determining the optimal number of visits $\totalNumberOfVisits^*$ when given the maximum number of visits possible $\maximumNumberOfVisits$. Doing so allows a network operator to find the optimum visit expenditures, visit lifetimes, and number of visits.

We can derive an equation for NPC when $\totalNumberOfVisits$ visits are made and $\visitLifetimeFunction{\visitNumber}{\visitExpenditure{\visitNumber}} = \linearLifetimeFunction{\visitNumber}$ when $\visitExpenditure{\visitNumber} \ge -\lifetimeFunctionYIntercept{\visitNumber}/\lifetimeFunctionSlope$. Using the fact that $\sum_{\visitNumber=1}^{\totalNumberOfVisits} \visitLifetimeFunction{\visitNumber}{\visitExpenditure{\visitNumber}} = \operationalLifetime$, and letting $\expenditureVector = [ \visitExpenditure{1} \ \visitExpenditure{2} \ \ldots \ \visitExpenditure{\totalNumberOfVisits}]$, we can re-write the NPC \eref{eq:NPC-with-lifetime} when $\totalNumberOfVisits \ge 2$ as
\begin{IEEEeqnarray*}{rCl}
\npcFunction = && \frac{\operationalLifetime - \lifetimeFunctionYIntercept{1} - \sum_{\visitNumber = 2}^{\totalNumberOfVisits}\left(\lifetimeFunctionSlope\visitExpenditure{\visitNumber}+\lifetimeFunctionYIntercept{\visitNumber}\right)}{\lifetimeFunctionSlope} 
\\ & + & {\sum_{\visitNumber = 2}^{\totalNumberOfVisits}\frac{\visitExpenditure{\visitNumber}}{(1+\interestRate)^{\operationalLifetime-\sum_{\visitExpenditureIndex = \visitNumber}^{\totalNumberOfVisits}\left(\linearLifetimeFunction{\visitExpenditureIndex}\right)}}}
\\ & + & \sum_{\unscheduledPaymentNumber=1}^{\lfloor \operationalLifetime / \mtbf \rfloor} \frac{\unscheduledPayment{}}{(1+\interestRate)^{ \unscheduledPaymentNumber \mtbf}}.
\end{IEEEeqnarray*}
Note that the first term represents the initial visit expenditure $\visitExpenditure{1}$. When $\totalNumberOfVisits=1$, the initial expenditure $\visitExpenditure{1}$ is set to the value that ensures an operational lifetime $\operationalLifetime$.

We can find the visit expenditure vector $\expenditureVector = [\visitExpenditure{2} \ \ldots \ \visitExpenditure{\totalNumberOfVisits}]$ that minimizes the NPC $\npcFunction$ for $\totalNumberOfVisits \ge 2$ using the Lagrangian
$$
\lagrangian = \npcFunction - \sum_{\visitNumber=2}^\totalNumberOfVisits \slackVariable{\visitNumber} (\visitExpenditure{\visitNumber}+\lifetimeFunctionYIntercept{\visitNumber}/\lifetimeFunctionSlope),
$$
where $\slackVariableVector=[\slackVariable{2} \, \ldots \, \slackVariable{\totalNumberOfVisits}]$. When solving a minimization problem with inequality constraints, the slack variables $\slackVariable{\visitNumber}$ are optimized to satisfy the Karush-Kuhn-Tucker (KKT) conditions. The complementary slackness KKT condition implies that $\visitExpenditure{\visitNumber} = -\lifetimeFunctionYIntercept{\visitNumber}/\lifetimeFunctionSlope$ when $\slackVariable{\visitNumber} > 0$, and $\slackVariable{\visitNumber} = 0$ when $\visitExpenditure{\visitNumber} > -\lifetimeFunctionYIntercept{\visitNumber}/\lifetimeFunctionSlope$. 

We can use the gradient of the Lagrangian to derive an equation for finding $\visitExpenditure{\visitNumber+1}$ from $\visitExpenditure{\visitNumber}$. The stationarity KKT condition requires $\nabla \lagrangian=\zeroVector$, where $\nabla \lagrangian = [\frac{\partial \lagrangian}{\partial \visitExpenditure{1}} \ \ldots \ \frac{\partial \lagrangian}{\partial \visitExpenditure{\totalNumberOfVisits}} \frac{\partial \lagrangian}{\partial \slackVariable{1}} \ \ldots \ \frac{\partial \lagrangian}{\partial \slackVariable{\totalNumberOfVisits}}]$. Subtracting $\frac{\partial \lagrangian}{\partial\visitExpenditure{\visitNumber+1}} - \frac{\partial \lagrangian}{\partial \visitExpenditure{\visitNumber}}$ for any $\visitNumber \le \totalNumberOfVisits-1$, yields
$$
\visitExpenditure{\visitNumber+1} = \nextPaymentTermOne{\visitExpenditure{\visitNumber}} + \frac{\slackVariable{\visitNumber+1} - \slackVariable{\visitNumber}}{\lifetimeFunctionSlope \ln(1+\interestRate) (1+\interestRate)^{\operationalLifetime-\lifetimeFunctionSlope \visitExpenditure{\visitNumber+1}-\lifetimeFunctionYIntercept{\visitNumber+1}}},
$$
where
$$
\nextPaymentTermOne{\visitExpenditure{\visitNumber}} = \frac{ \left( 1 + \interestRate \right)^{\linearLifetimeFunction{\visitNumber}} - 1}{\lifetimeFunctionSlope\ln{\left(1+\interestRate\right)}}.
$$
The visit expenditure $\visitExpenditure{\visitNumber+1}$ can be found using
\begin{equation}
\label{eq:next-payment}
\visitExpenditure{\visitNumber+1} = 
\begin{cases}
\nextPaymentTermOne{\visitExpenditure{\visitNumber}} & \nextPaymentTermOne{\visitExpenditure{\visitNumber}} > -\lifetimeFunctionYIntercept{\visitNumber+1}/\lifetimeFunctionSlope \\
-\lifetimeFunctionYIntercept{\visitNumber+1}/\lifetimeFunctionSlope & \nextPaymentTermOne{\visitExpenditure{\visitNumber}} \le -\lifetimeFunctionYIntercept{\visitNumber+1}/\lifetimeFunctionSlope.
\end{cases}
\end{equation}
Keeping in mind that $-\lifetimeFunctionYIntercept{\visitNumber+1}/\lifetimeFunctionSlope \ge 0$, if $\nextPaymentTermOne{\visitExpenditure{\visitNumber}} > -\lifetimeFunctionYIntercept{\visitNumber+1}/\lifetimeFunctionSlope$, then $\visitExpenditure{\visitNumber} > -\lifetimeFunctionYIntercept{\visitNumber}/\lifetimeFunctionSlope$ so $\slackVariable{\visitNumber}=0$. As $\visitExpenditure{\visitNumber+1} \ge \nextPaymentTermOne{\visitExpenditure{\visitNumber}} > -\lifetimeFunctionYIntercept{\visitNumber+1}/\lifetimeFunctionSlope$, $\slackVariable{\visitNumber+1} = 0$ so $\visitExpenditure{\visitNumber+1} = \nextPaymentTermOne{\visitExpenditure{\visitNumber}}$. When $\nextPaymentTermOne{\visitExpenditure{\visitNumber}} \le -\lifetimeFunctionYIntercept{\visitNumber+1}/\lifetimeFunctionSlope$, then $\slackVariable{\visitNumber+1}>\slackVariable{\visitNumber}$ to ensure that $\visitExpenditure{\visitNumber+1} \ge -\lifetimeFunctionYIntercept{\visitNumber+1}/\lifetimeFunctionSlope$; to satisfy the complementary slackness KKT condition, making $\slackVariable{\visitNumber+1}>\slackVariable{\visitNumber}\ge0$ forces $\visitExpenditure{\visitNumber+1} = -\lifetimeFunctionYIntercept{\visitNumber+1}/\lifetimeFunctionSlope$.

The KKT conditions are maintained by \eref{eq:next-payment}, meaning it can be used to find a local optimum solution. By ensuring that $\slackVariable{\visitNumber} = 0$ when $\visitExpenditure{\visitNumber} > -\lifetimeFunctionYIntercept{\visitNumber}/\lifetimeFunctionSlope$, and $\visitExpenditure{\visitNumber} = -\lifetimeFunctionYIntercept{\visitNumber}/\lifetimeFunctionSlope$ when $\slackVariable{\visitNumber} > 0$, it satisfies the complimentary slackness condition. The primal and dual feasibility conditions are satisfied by ensuring that $\visitExpenditure{\visitNumber} \ge -\lifetimeFunctionYIntercept{\visitNumber}/\lifetimeFunctionSlope$ and $\slackVariable{\visitNumber} \ge 0$ respectively. To derive \eref{eq:next-payment}, the stationarity condition $\nabla \lagrangian=\zeroVector$ was assumed.

When $\nextPaymentTermOne{\visitExpenditure{\visitNumber}} \le -\lifetimeFunctionYIntercept{\visitNumber+1}/\lifetimeFunctionSlope$ and $\visitExpenditure{\visitNumber+1} = -\lifetimeFunctionYIntercept{\visitNumber+1}/\lifetimeFunctionSlope$, note that no energy is being added to the network. Visit $\visitNumber+1$, and every subsequent visit, will have a visit lifetime of 0. If this occurs, the chosen $\totalNumberOfVisits$ value is too high and is not the optimal $\totalNumberOfVisits^*$ value. 

Before discussing our method of using \eref{eq:next-payment} to minimize $\npcFunction$, we first prove that the local minimum of $\npcFunction$ is also the global minimum by showing that $\npcFunction$ is convex when $\totalNumberOfVisits \ge 2$. We will do so by letting
$$
\npcExpenditureFunction = \frac{\visitExpenditure{\visitNumber}}{(1+\interestRate)^{\operationalLifetime-\sum_{\visitExpenditureIndex = \visitNumber}^{\totalNumberOfVisits}\left(\linearLifetimeFunction{\visitExpenditureIndex}\right)}},
$$
and proving that $\npcExpenditureFunction$ is a convex function with respect to the visit expenditures $\visitExpenditure{\visitNumber}$. If $\npcExpenditureFunction$ is convex, then $\npcFunction$ is the sum of convex functions and is therefore itself convex. The second partial derivative of $\npcExpenditureFunction$ is
$$
\frac{\partial^2 \npcExpenditureFunction}{\partial \visitExpenditure{i} \partial \visitExpenditure{j}} =
\begin{cases}
\frac{2\lifetimeFunctionSlope\ln(1+\interestRate) + \lifetimeFunctionSlope^2\ln^2(1+\interestRate)\visitExpenditure{\visitNumber}}{(1+\interestRate)^{\operationalLifetime-\sum_{\visitExpenditureIndex = \visitNumber}^{\totalNumberOfVisits}\left(\linearLifetimeFunction{\visitExpenditureIndex}\right)}} & i=\visitNumber \land j=\visitNumber \\
\frac{\lifetimeFunctionSlope\ln(1+\interestRate) + \lifetimeFunctionSlope^2\ln^2(1+\interestRate)\visitExpenditure{\visitNumber}}{(1+\interestRate)^{\operationalLifetime-\sum_{\visitExpenditureIndex = \visitNumber}^{\totalNumberOfVisits}\left(\linearLifetimeFunction{\visitExpenditureIndex}\right)}} & \parbox{100pt}{$\left(i=\visitNumber \land j > \visitNumber\right)$ \\ \hphantom{} $\, \lor \left(j=\visitNumber \land i > \visitNumber\right) $}\\
\frac{\lifetimeFunctionSlope^2\ln^2(1+\interestRate)\visitExpenditure{\visitNumber}}{(1+\interestRate)^{\operationalLifetime-\sum_{\visitExpenditureIndex = \visitNumber}^{\totalNumberOfVisits}\left(\linearLifetimeFunction{\visitExpenditureIndex}\right)}} & i>\visitNumber \land j>\visitNumber \\
0 & i < \visitNumber \lor j < \visitNumber. \\
\end{cases}
$$
By definition $\visitExpenditure{\visitNumber} \ge -\lifetimeFunctionYIntercept{\visitNumber}/\lifetimeFunctionSlope \ge 0$, ensuring that the elements of the Hessian $\hessian$ of $\npcExpenditureFunction$ are all non-negative, and that the elements of the visit expenditure vector $\expenditureVector$ are non-negative real numbers. This means $\expenditureVector \hessian \expenditureVector^{\mathsf{T}} \ge 0$, so $\hessian$ is a positive semi-definite matrix. As $\hessian$ is positive semi-definite, $\npcExpenditureFunction$ is a convex function, so $\npcFunction$ is the sum of convex functions and is therefore convex itself. The local optimum found when minimizing NPC for $\totalNumberOfVisits$ visits is consequently the global optimum.

\aref{alg:op} can be applied to find the global minimum of $\npcFunction$. It determines the optimal value of $\visitExpenditure{2}$, then applies \eref{eq:next-payment} to find the remaining visit payments. The function \textsc{FindRoot}$(f(x))$ finds $x>0$ such that $f(x)=0$. The derivative $\frac{\partial \lagrangian}{\partial \visitExpenditure{2}}$ of the NPC's Lagrangian $\lagrangian$ is expressed in terms of $\visitExpenditure{2}$ and $\totalNumberOfVisits$ by \textsc{dLagrangian}$(\visitExpenditure{2},\totalNumberOfVisits)$. As the local minimum of $\npcFunction$ is its global minimum, calling \textsc{FindRoot}(\textsc{dLagrangian}$(\visitExpenditure{2},\totalNumberOfVisits)$) finds the value of $\visitExpenditure{2}$ that minimizes $\npcFunction$. The function \textsc{nextP}$(\visitExpenditure{\visitNumber},\totalNumberOfVisits)$ uses $\visitExpenditure{2}$ and \eref{eq:next-payment} to calculate the visit payments $\visitExpenditure{3},\visitExpenditure{4},\ldots,\visitExpenditure{\totalNumberOfVisits}$.

\begin{algorithm}
\caption{Optimizing the visit expenditures for a fixed number of visits $\totalNumberOfVisits$.}
\label{alg:op} 
\begin{algorithmic}
\State // Returns the visit expenditure vector $\expenditureVector$ that achieves the
\State // optimal NPC if $\totalNumberOfVisits \geq 2$ visits are performed.
 \Function{OptimalPayments}{$\totalNumberOfVisits$}
 \State // Find the optimal value of $\visitExpenditure{2}$ 
 \Let{$p_2$}{} \Call{FindRoot}{\Call{dLagrangian}{$\visitExpenditure{2}$,$\totalNumberOfVisits$}}
  \State // Find the remaining payment values
\For{$\visitNumber \gets 2 \text{ to } (\totalNumberOfVisits-1)$} 
\Let{$\visitExpenditure{\visitNumber+1}$}{} \Call{nextP}{$\visitExpenditure{\visitNumber}$,$\totalNumberOfVisits$}
\EndFor
\Let{$\visitExpenditure{1}$}{$\frac{\operationalLifetime - \lifetimeFunctionYIntercept{1} - \sum_{\visitNumber = 2}^{\totalNumberOfVisits}\left(\lifetimeFunctionSlope \visitExpenditure{\visitNumber}+\lifetimeFunctionYIntercept{\visitNumber}\right)}{\lifetimeFunctionSlope}$}
 \State \Return $\expenditureVector$
 \EndFunction
\end{algorithmic}
\end{algorithm}

We use \aref{alg:onpc} to determine the number of visits $\totalNumberOfVisits$ that optimize the NPC. It finds the NPC for every $\totalNumberOfVisits \in [1,\maximumNumberOfVisits]$ using \aref{alg:op}, and chooses the values $\totalNumberOfVisits^*$ and $\expenditureVector^*$ that minimize the NPC. The function \textsc{NPC}$(\expenditureVector,\totalNumberOfVisits)$ uses \eref{eq:NPC-with-lifetime} to calculate the NPC. \aref{alg:onpc} is bounded by $\bigO{\maximumNumberOfVisits^3}$ operations. In order to express $\visitExpenditure{k}$ in terms of $\visitExpenditure{2}$, \eref{eq:next-payment} must be applied $\visitNumber-2$ times. The number of operations in the equation for $\frac{\partial \lagrangian}{\partial \visitExpenditure{2}}$, after all $\visitExpenditure{\visitNumber}$ values are expressed in terms of $\visitExpenditure{2}$, are bounded by $\bigO{\totalNumberOfVisits^2}$. Expressing $\totalNumberOfVisits-2$ values in terms of $\visitExpenditure{2}$ requires applying \eref{eq:next-payment} a total of $(\totalNumberOfVisits-1)(\totalNumberOfVisits-2)/2$ times, leading to the $\bigO{\totalNumberOfVisits^2}$ bound. Once $\visitExpenditure{2}$ is found, the number of operations to find the remaining $\totalNumberOfVisits-1$ visit expenditure values are bounded by $\bigO{\totalNumberOfVisits}$. As $\totalNumberOfVisits \in [1,\maximumNumberOfVisits]$, we have to repeat these steps $\maximumNumberOfVisits$ times, meaning that finding the optimal number of visits $\totalNumberOfVisits^*$ and the corresponding optimal NPC $\npcFunctionOpt$ is an $\bigO{\maximumNumberOfVisits^3}$ algorithm.

\begin{algorithm}
\caption{Finding the optimal number of visits and NPC value for up to $\maximumNumberOfVisits$ visits.}
\label{alg:onpc}  
\begin{algorithmic}
 \Function{OptimalNumberOfVisits}{$\maximumNumberOfVisits$}
 \Let{$\npcVariable^*$}{$\frac{\operationalLifetime-\lifetimeFunctionYIntercept{1}}{\lifetimeFunctionSlope}$}
 \Let{$\totalNumberOfVisits^*$}{$1$}
 \State // Try every value of $\totalNumberOfVisits$
 \For{$\totalNumberOfVisits \gets 2 \text{ to } \maximumNumberOfVisits$}
 \Let{$\expenditureVector$}{} \Call{OptimalPayments}{$\totalNumberOfVisits$}
 \Let{$\npcVariable$}{} \Call{NPC}{$\expenditureVector$,$\totalNumberOfVisits$}
 \If{$\npcVariable<\npcVariable^*$}
 	\Let{$\npcVariable^*$}{\npcVariable}
 	 \Let{$\totalNumberOfVisits^*$}{$\totalNumberOfVisits$}
 \EndIf
 \EndFor
 \State \Return $\totalNumberOfVisits^*$
 \EndFunction
\end{algorithmic}
\end{algorithm}

\subsection{Equal Visit Lifetime Approximation}
\label{ssec:equal-visit-lifetime-approximation}

Equally spacing the visits apart can be a good rule of thumb when the lifetime function \visitLifetimeFunction{\visitNumber}{\visitExpenditure{\visitNumber}} is linear as in \sref{ssec:npc-minimization-with-linear-relationship}, and $|\lifetimeFunctionYIntercept{\visitNumber}| << 1, \forall \visitNumber \in [2,\totalNumberOfVisits]$. In the previous sections the visit lifetimes could be non-uniform; they could vary for each visit $\visitNumber$ depending on the visit expenditure $\visitExpenditure{\visitNumber}$. In this section we prove that visit lifetimes $\visitLifetime{\visitNumber}$ are approximately equal as long as $\interestRate << 1$ and $|\lifetimeFunctionYIntercept{\visitNumber}| << 1, \forall \visitNumber \in [2,\totalNumberOfVisits]$.

We assume that the value of $\totalNumberOfVisits$ is chosen such that, $\forall \visitNumber \in [1,\totalNumberOfVisits]$, $\nextPaymentTermOne{\visitExpenditure{\visitNumber}} > -\lifetimeFunctionYIntercept{\visitNumber+1}/\lifetimeFunctionSlope$, so $\visitExpenditure{\visitNumber+1} = \nextPaymentTermOne{\visitExpenditure{\visitNumber}}$. We can use \eref{eq:next-payment} to express the lifetime $\visitLifetime{\visitNumber+1}$ of visit $\visitNumber+1$ in terms of the visit expenditure $\visitExpenditure{\visitNumber}$ of visit $\visitNumber$ through
$$
\visitLifetime{\visitNumber+1} = \frac{ \left( 1 + \interestRate \right)^{\linearLifetimeFunction{\visitNumber}} - 1}{\ln{\left(1+\interestRate\right)}} + \lifetimeFunctionYIntercept{\visitNumber+1}.
$$ 
We can write $(1+\interestRate)^{m \visitExpenditure{\visitNumber}+\lifetimeFunctionYIntercept{\visitNumber}} = \exp[\ln(1+\interestRate)(\lifetimeFunctionSlope\visitExpenditure{\visitNumber}+\lifetimeFunctionYIntercept{\visitNumber})]$, and as long as $\ln(1+\interestRate)(\lifetimeFunctionSlope\visitExpenditure{\visitNumber}+\lifetimeFunctionYIntercept{\visitNumber}) << 1$, $\exp[\ln(1+\interestRate)(\lifetimeFunctionSlope\visitExpenditure{\visitNumber}+\lifetimeFunctionYIntercept{\visitNumber})] \approx 1 + \ln(1+\interestRate)(\lifetimeFunctionSlope\visitExpenditure{\visitNumber}+\lifetimeFunctionYIntercept{\visitNumber})$, so after simplification
$$
\visitLifetime{\visitNumber+1} \approx \lifetimeFunctionSlope\visitExpenditure{\visitNumber}+\lifetimeFunctionYIntercept{\visitNumber}+\lifetimeFunctionYIntercept{\visitNumber+1},
$$
which can be simplified to
\begin{equation}
\label{eq:next-payment-linear}
\visitLifetime{\visitNumber+1} \approx \visitLifetime{\visitNumber}+\lifetimeFunctionYIntercept{\visitNumber+1}.
\end{equation}

As long as $\interestRate << 1$ and $|\lifetimeFunctionYIntercept{\visitNumber}| << 1, \forall \visitNumber \in [2,\totalNumberOfVisits]$, evenly spacing the lifetimes is a near-optimal approach. If $\interestRate << 1$ and $|\lifetimeFunctionYIntercept{\visitNumber}| << 1$, then it is likely that $\ln(1+\interestRate)(\lifetimeFunctionSlope\visitExpenditure{\visitNumber}+\lifetimeFunctionYIntercept{\visitNumber}) << 1$, which is required by the approximation.  The magnitude of $\lifetimeFunctionYIntercept{1}$ does not have to be small; in fact, chances are it will not be small due to the node hardware costs of the initial visit. We call this the \emph{Equal Visit Lifetime (EVL)} approximation.

\subsection{Minimizing the Initial Cost of Node Hardware and the Approximate NPC of Unscheduled Maintenance}
\label{ssec:minimizing-unscheduled-npc}

To reduce the NPC of unscheduled payments, a network operator could purchase more robust nodes or add redundant nodes in stand-by to node locations in order to increase the network's reliability. This, however, increases the initial cost of node hardware. By increasing the initial cost of node hardware $\nodeHardwareExpenditure{1}$, the network operator can increase the spacing between the failure times $\unscheduledPaymentTime{1} \ldots \unscheduledPaymentTime{\numberOfFailures}$ of the network and reduce the number of failures $\numberOfFailures$ that occur over the network's operational lifetime. The NPC \eref{eq:NPC} will decrease if the reduction in $\sum_{\unscheduledPaymentNumber=1}^{\numberOfFailures} \frac{\unscheduledPayment{\unscheduledPaymentNumber}}{(1+\interestRate)^{\unscheduledPaymentTime{\unscheduledPaymentNumber}}}$ exceeds the increase in $\nodeHardwareExpenditure{1}$. When $\nodeHardwareExpenditure{1}$ and $\sum_{\unscheduledPaymentNumber=1}^{\numberOfFailures} \frac{\unscheduledPayment{\unscheduledPaymentNumber}}{(1+\interestRate)^{\unscheduledPaymentTime{\unscheduledPaymentNumber}}}$ have been found, the NPC can be minimized using \eref{eq:npc-minimization} if the lifetime function is non-linear, or using the methods provided in \sref{ssec:npc-minimization-with-linear-relationship} and \sref{ssec:equal-visit-lifetime-approximation} if the lifetime function is linear. This section provides a method for balancing $\nodeHardwareExpenditure{1}$, $\unscheduledPaymentTime{\unscheduledPaymentNumber}$, and $\numberOfFailures$ to minimize the cost of node hardware and the approximate NPC of unscheduled payments in \eref{eq:NPC-with-lifetime}, where $\mtbf$ is the network's MTBF, $\unscheduledPaymentTime{\unscheduledPaymentNumber} = \unscheduledPaymentNumber \mtbf$, and $\numberOfFailures=\lfloor \operationalLifetime/\mtbf \rfloor$.

We assume that the network operator is given a set of choices $\choiceSet$, with each choice $i$ costing $\nodeCost{i}$ financial units per node location while having a failure rate of $\failureRate{i}$ per node location. Choices could represent different node hardware, each with their own cost and failure rate. Alternatively, the choices could also include different levels of redundancy. Assuming Poisson failures, putting $\numberOfRedundantNodes$ nodes at a location would reduce the failure rate by a factor of $\numberOfRedundantNodes$ while increasing the cost by a factor of $\numberOfRedundantNodes$.

We assume that node hardware is only added during the initial deployment, that all nodes at a single node location failing leads to a network failure, that the network operator knows the number of nodes $\numberOfNodes$ that are required by the network, that every node has the same type of node hardware, and that every node location is given the same number of redundant nodes in stand-by. The approximate cost of each choice with respect to the initial payment and unscheduled payments can then be written as
\begin{equation}
\label{eq:hardware-and-unscheduled-payment-npc}
\numberOfNodes \nodeCost{i} + \sum_{n=1}^{\lfloor \operationalLifetime \numberOfNodes \failureRate{i} \rfloor}\frac{\unscheduledPayment{} + \nodeCost{i}}{(1+\interestRate)^{\frac{n}{\numberOfNodes \failureRate{i}}}},
\end{equation}
where $\unscheduledPayment{}$ is the cost of performing the repair. To find the optimum choice $i^*$, the operator can try every choice and pick the one that minimizes \eref{eq:hardware-and-unscheduled-payment-npc}. The number of operations required are bounded by $\bigO{\sum_{i \in \choiceSet} \operationalLifetime \numberOfNodes \failureRate{i}}$.

\section{Numerical Results}
\label{sec:numerical-results}

In this section we provide numerical results using the frameworks from \sref{sec:visit-lifetime-framework} and \sref{sec:multiple-deployment-optimization}. \sref{ssec:individual-simulation-parameters} presents node and network assumptions based on a gas monitoring scenario. We analyse the effectiveness of our method for balancing initial node hardware costs with the approximate NPC of unscheduled payments in \sref{ssec:approx-hardware-unscheduled-payment-npc-analysis}. In \sref{ssec:comparing-payment-strategies} we demonstrate that NPC minimization is effective at reducing costs and that equally spacing visits is a near-optimal strategy. \sref{ssec:analysis-of-npc-minimization-parameters} demonstrates that NPC minimization is most effective when node hardware and labor expenditures are low and a large portion of the budget is spent on energy. \sref{ssec:npc-cost-breakdown} examines how adjusting the number of visits $\totalNumberOfVisits$, the operational lifetime $\operationalLifetime$, the interest rate $\interestRate$, the network's MTBF $\mtbf$, and other parameters related to the cost of node hardware and labor affects the portion of the NPC allocated to node hardware, failure repair, energy, and labor.

The \emph{percent savings} is used in this section to measure the effectiveness of NPC minimization when compared to a network design based on a single payment at the start of the network lifetime. The minimized NPC value is the cost $\npcValueFunction{[\visitExpenditure{1} \ldots \visitExpenditure{\totalNumberOfVisits}]}$ of running a network for $\operationalLifetime$ years by performing NPC minimization over $\totalNumberOfVisits$ visits. The single payment benchmark is the cost, $\singleExpenditureBenchmark$, of providing each node with enough energy to last $\operationalLifetime$ years at the time of initial deployment; note that $\singleExpenditureBenchmark$ is the NPC when $\totalNumberOfVisits=1$. The percent savings is defined as the percent difference between these two values such that
$$
\frac{|\singleExpenditureBenchmark - \npcValueFunction{[\visitExpenditure{1} \ldots \visitExpenditure{\totalNumberOfVisits}]}|}{\singleExpenditureBenchmark}.
$$

In this section we use the dollar symbol \$ to represent a financial unit. The NPC minimization framework does not assume a specific currency; any currency can been used as long as it is consistent for all costs. Prices given for node hardware, energy, and labor reflect their value in United States dollars at the time this paper was written.

\subsection{Node and Network Assumptions}
\label{ssec:individual-simulation-parameters}

\begin{figure}
\centering
\includegraphics[scale=0.17]{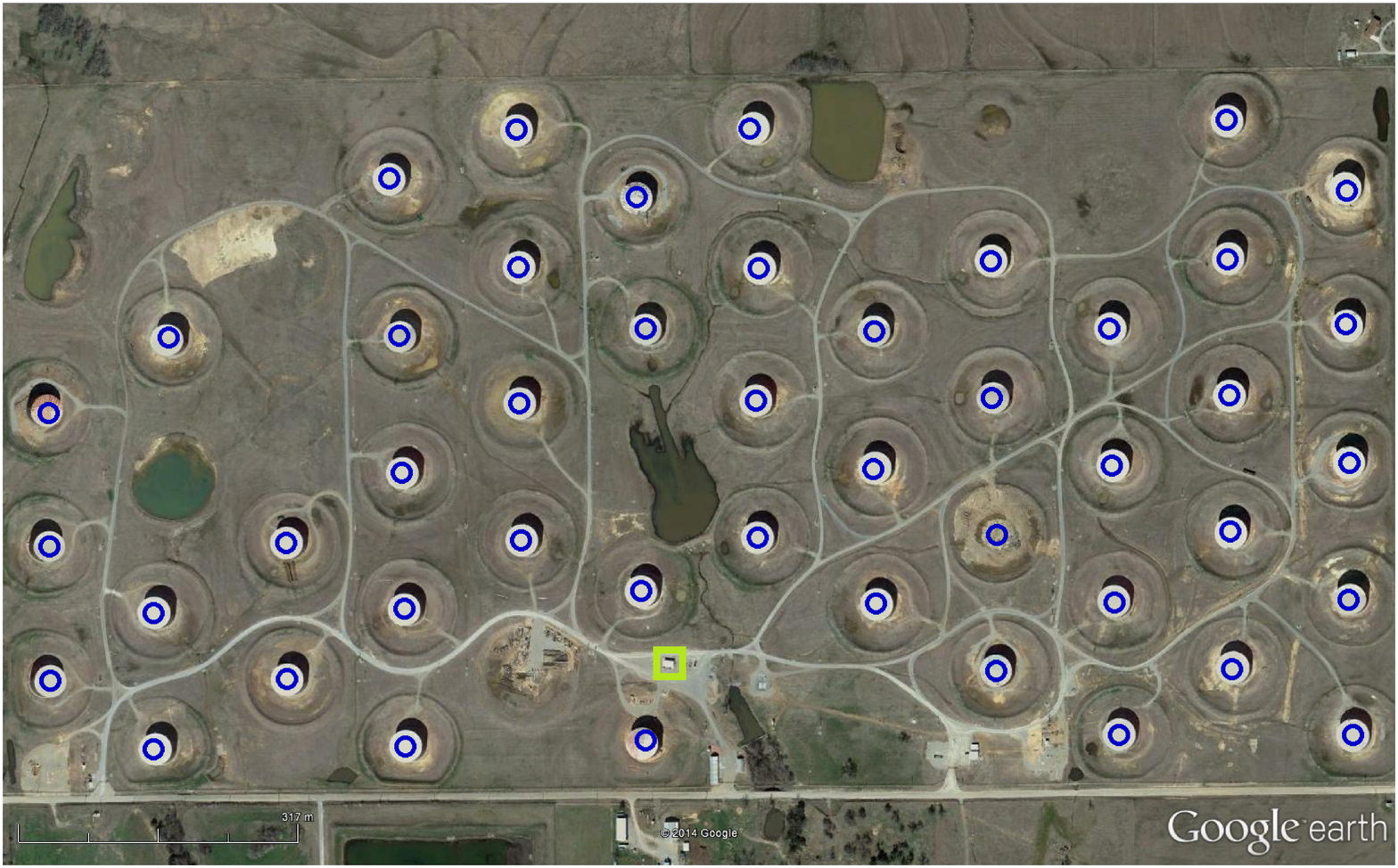}
\caption{A possible area for monitoring in our scenario. Map data and image \copyright 2014 Google.}
\label{fig:tanks}
\end{figure}

We consider applying the network and cost models from \sref{sec:network-and-cost-models} to a gas monitoring scenario, where a company could install a WSN to continuously monitor the concentration of hazardous gas in the air at a storage site. The satellite image in \fref{fig:tanks} provides an aerial view of one area where such a network could be built. When finding the parameters for our numerical results, we assume that sensor nodes are the circles positioned on top of the tanks in the picture. The sink node is located in the middle of the square. To connect disjoint sensors to the sink, we populate the candidate relay location set, $\candidateRelayLocationSet{}$, with locations spaced along a minimum Steiner tree generated by the GeoSteiner \cite{Warme2000} tool using the process described in \cite{Brazil2009}.

In our scenario, we assume that the relay and sensor nodes are similar to Iris motes \cite{iris} and that TGS 825 gas sensors \cite{tgs825} are used. The gas sensor is heated for about \SI{60}{\second} prior to each measurement, meaning \SI{39.6}{\joule} are consumed per measurement; the measurements occur every 5 minutes. The energy consumed by the RF230 transceiver and ATmega128l microcontroller present on Iris motes while receiving and transmitting data are taken from their datasheets \cite{rf230} and \cite{Atmega128l} respectively. We also assume they are connected to a \SI{3}{\volt} source, that the RF230's data rate is \SI{250}{\kilo bit/\second}, and that nodes transmit data in \SI{64}{bit} packets. Note that in general, sensor nodes measuring gas concentration require a large amount of energy due to the heating elements in the gas sensors, and consume significantly more energy than the relay nodes in the network. The results in this section, however, do not rely on sensor nodes consuming more energy than relay nodes. Similar results would be found in a scenario where relays consume more energy than sensors, such as a video surveillance network where relays may have to forward large amounts of data frequently, as long as the network consumes power at a similar rate.

We minimize the power $\transmitPowerFunction{i}{j}$ consumed by the RF230 on node $i$ by adjusting its transmit power when sending data to a node $j$ that is $\distance{i}{j}$\,m away. The transmit signal power $\signalTransmitPower{i}{j}$\,W required at node $i$ to ensure a receive signal power $\signalReceivePower$\,W at a destination node $j$ is calculated using the Friis equation
\begin{equation}
\label{eq:friis}
\signalTransmitPower{i}{j} = \frac{\signalReceivePower(4 \pi \distance{0}{})^2}{\antennaGain{i}\antennaGain{j}\wavelength^2}\left(\frac{\distance{i}{j}}{\distance{0}{}}\right)^\pathlossExponent,
\end{equation}
where $\distance{0}{}=1$\,m is the distance from the antenna to the edge of the near field, $\antennaGain{i}=\antennaGain{j}=1.5$ is the antenna gain of a dipole antenna, $\wavelength=$\ \SI{125}{\milli \meter} is the wavelength of the signal, and $\pathlossExponent=$\ 4 is the path loss exponent. We set the receive signal power $\signalReceivePower=\ $\SI{-101}{dBm}, which is the receiver sensitivity of the Iris mote. We calculate the 16 discrete RF230 transmit signal powers with a quadratic interpolation of the transmit current consumption values for different transmit signal powers given in its datasheet, which are then used with \eref{eq:friis} to determine the power consumed by a node $i$ when transmitting to another node $j$ that is $\distance{i}{j}$\,m away.

As stated in \sref{ssec:optimization-problems}, we assume that the node hardware cost consists only of purchasing the nodes for initial deployment.  We adjust the cost of a single node, $\nodeCost{}$, between \SI{10}[\$] and \SI{100}[\$]; such a range covers cases where an operator builds nodes independently and where an operator purchases them from a manufacturer. By running \eref{eq:mip-lifetime-optimization} on the network in \fref{fig:tanks}, we find that about 150 nodes in total must be purchased.

In \sref{ssec:optimization-problems} we discuss that the portion of an expenditure dedicated to energy depends on the cost of each Joule and the amount of Joules required by the network. The cost of a Joule $\energyCost$ when using alkaline D-Cell batteries is approximately \SI{20}{\micro \$/\joule}, but is \SI{66}{\micro \$/\joule} for Lithium D-Cell batteries, which are lighter, have a higher capacity, and are better suited for extreme climates. The network described in \sref{ssec:individual-simulation-parameters} consumes energy at a rate $\powerConsumption$ of approximately \SI{6.2}{\watt} while running.

We assume that the price of labor $\laborExpenditure{\visitNumber}$ for each visit $\visitNumber$ does not change over time; that is $\forall \visitNumber \in [1,\totalNumberOfVisits]: \laborExpenditure{\visitNumber} = \laborCost$ in dollars. To find the labor cost $\laborCost$, we ran the Traveling Salesman Problem \cite{Gutin2002} on the network in \fref{fig:tanks}; when assuming a walking speed of about \SI{1.4}{\meter/\second} \cite{Browning2006}, it takes approximately 2 hours for someone to visit all of the nodes. With a wage of \SI{20}{\$/\hour}, and assuming \SI{60}{\second} per node is taken for swapping batteries, we assume $\laborCost$ is minimum \$140. The labor cost $\laborCost$ would likely be higher for networks with nodes that are difficult to reach, networks that are in remote areas, or networks in areas that are difficult to traverse. We assume that a visit could cost up to \$1000 in such scenarios, due to higher wages, extra time spent at the site, and transportation costs in order to reach the site.

We assume that all nodes have identical failure rates. While we were unable to find failure rate data on commonly used motes, such as the Iris mote, we were able to obtain wired gas monitor failure rates from Draeger \cite{Draeger2012}. The DraegerSense IR has one of the best failure rates of about \SI{0.5}{\micro failures/\hour}. While Draeger sensors are not wireless, the CC2420 transceiver has a failure rate in the order of \SI{1.9}{\nano failures/\hour} \cite{cc2420quality}. We assume that a high-quality node with a gas sensor could have a slightly higher failure rate of \SI{0.75}{\micro failures/\hour}.  We assume lower-quality nodes have failure rates of \SI{10}{\micro failures/\hour}, higher than that of Draeger Polytron 7000 gas monitors with electrochemical sensors. The high-quality and low-quality node hardware failure rates correspond to network MTBF $\mtbf$ of about \SI{1}{year} and \SI{1}{month} respectively.

We set the cost of each repair to \$1000. While it takes less time to replace a single sensor than to restore energy to all $\numberOfNodes=150$\,nodes, we cannot predict when failures occur, so unlike replenishing energy, there will be downtime between the beginning of the failure and its repair. The repair cost value accounts for the network being unable to perform its function during this downtime, as well as the cost of having some type of emergency responder available to repair the network.

We placed limits on the operational lifetime of the network and the interest rate when minimizing NPC. Operational lifetime, $\operationalLifetime$, ranges from 1 to 10 years. Based on an average stock market rate of return of 0.083 over the last 114 years in developed countries \cite{Dimson2014}, the interest rate is assumed to vary between 0.01 and 0.1 to capture below and above average rates of return.

In the following sections, results are generated by assuming a default set of parameters. We assume the cost $\nodeCost{}$ of each node is \SI{10}[\$], 150 nodes are present in the network, the cost $\energyCost$ of a single Joule is \SI{20}{\micro \$/\joule}, the network consumes energy at a rate $\powerConsumption$ of \SI{6.2}{\watt}, the cost $\laborCost$ of labor is \SI{1000}{\$/visit}, the MTBF $\mtbf$ is \SI{1}{year}, and the cost $\unscheduledPayment{}$ of each failure repair is \$1000. By default, the operational lifetime $\operationalLifetime$ is 10 years, and the interest rate $\interestRate$ is 0.1. All results in the following sections are generated using these default values unless stated otherwise.

\subsection{Analysis of Minimizing the Initial Cost of Node Hardware and the Approximate NPC of Unscheduled Maintenance }
\label{ssec:approx-hardware-unscheduled-payment-npc-analysis}

In this section we demonstrate how the initial cost of node hardware and the approximate NPC of unscheduled payments can be minimized using the method described in \sref{ssec:minimizing-unscheduled-npc}. We assume that we have cheap, low quality nodes, and we want to find the optimal number of redundant nodes to place in stand-by at every node location in the network, assuming that the network fails if all the nodes at any one node location fail. We demonstrate that the method in \sref{ssec:minimizing-unscheduled-npc} can significantly reduce the approximate NPC of initial hardware and unscheduled payments.

Our choice set $\choiceSet$ consists of placing 1 to 10 nodes at each node location, where one node per location is active and the rest are in stand-by. Each node costs \SI{10}[\$] and has a failure rate of \SI{10}{\micro failures/\hour}. We assume Poisson failures, so putting $\numberOfRedundantNodes$ nodes at a location reduces the failure rate by a factor of $\numberOfRedundantNodes$ and increases the cost by a factor of $\numberOfRedundantNodes$.

\begin{figure}
  \centering
  \includegraphics[width=0.33\textwidth]{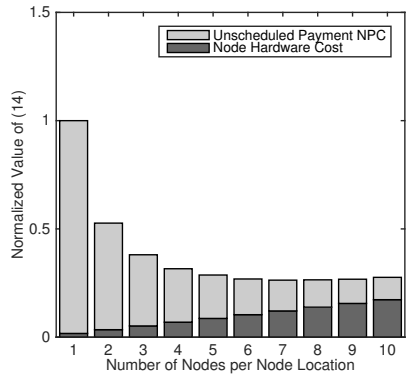}
  \caption{Normalized value of \eref{eq:hardware-and-unscheduled-payment-npc} for various numbers of nodes placed at each node location in the network.}
  \label{fig:redundancy}
\end{figure}

\fref{fig:redundancy} shows the value of \eref{eq:hardware-and-unscheduled-payment-npc} for each number of nodes per node location tested, normalized to only deploying one node at each location. The lower bar represents the value of node hardware, while the upper bar is the approximate NPC of unscheduled payments.

We can see in \fref{fig:redundancy} that the method in \sref{ssec:minimizing-unscheduled-npc} can significantly reduce the approximate NPC of node hardware and unscheduled payments. Node hardware costs increase linearly, but for small numbers of redundant nodes the approximate NPC of unscheduled payments decreases rapidly. After a certain point, for example in \fref{fig:redundancy} after 7 nodes have been placed at each node location, savings from additional redundancy no longer outpace additional hardware costs.

\subsection{Comparison of Optimization Strategies}
\label{ssec:comparing-payment-strategies}

\begin{figure}
  \centering
  \includegraphics[width=0.33\textwidth]{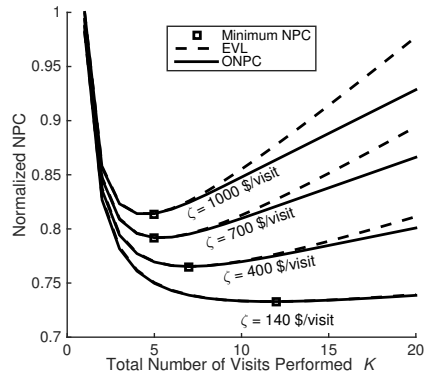}
  \caption{NPC for various numbers of visits when the cost of labor $\laborCost$ varies between \SI{140}{\$/visit} and \SI{1000}{\$/visit}.}
  \label{fig:npc-m500}
\end{figure}

In this section we compare the Equal Visit Lifetime (EVL) approximation to finding the Optimal NPC (ONPC). The lifetime function is linear, so the ONPC is found using the approach described in \sref{ssec:npc-minimization-with-linear-relationship}. \fref{fig:npc-m500} shows how the NPC found with both methods depends on the total number of visits $\totalNumberOfVisits$ and the labor cost $\laborCost$ in \$/visit. NPC is normalized to the cost of a network with an operational lifetime of $\operationalLifetime=10$ years when $\totalNumberOfVisits=1$ and $\laborCost=$\SI{1000}{\$/visit}. The value of $\laborCost$ written underneath each pair of EVL and ONPC lines was used to generate that pair of lines. The minimum ONPC value over all $\totalNumberOfVisits$ for a given value of $\laborCost$ is indicated by the squares.

The squares in \fref{fig:npc-m500} show that as the labor cost $\laborCost$ decreases, the minimum NPC over all numbers of visits $\totalNumberOfVisits$ decreases, while the optimal number of visits required to achieve that NPC increases. Decreasing the labor cost means that more visits can be performed before the cost of performing labor exceeds the savings of NPC minimization. More visits provide additional opportunities to exploit interest rates to purchase cheaper energy, reducing the money spent on energy to achieve a given operational lifetime $\operationalLifetime$, lowering the overall NPC of the network.

The rate of change of the EVL and ONPC lines are greater for higher values of $\laborCost$. The higher the labor cost, the greater the total cost of labor for a given number of visits $\totalNumberOfVisits$, and the greater the rate of change of the lines generated with that labor cost. When $\totalNumberOfVisits<5$ visits, NPC decreases rapidly with increasing $\totalNumberOfVisits$ because additional visits provide more opportunities to use interest rates to reduce the cost of energy. Eventually increasing the number of visits becomes ineffective at reducing energy costs. For instance, the slopes of the EVL and ONPC lines are positive when $\totalNumberOfVisits>5$ visits and $\laborCost=$\SI{1000}{\$/visit}, and the slopes of the EVL and ONPC lines are flat when $\totalNumberOfVisits>10$ visits and $\laborCost=$\SI{140}{\$/visit}.

The optimal NPCs calculated by the EVL approximation are within 0.01\% of the NPCs calculated using ONPC. For large values of $\totalNumberOfVisits$, however, the NPC found by EVL deviates from that of ONPC when $\laborCost$ is high, as shown in \fref{fig:npc-m500}. In \sref{ssec:equal-visit-lifetime-approximation} we assume that $|\lifetimeFunctionYIntercept{\visitNumber}|<<1$; however, the $|\lifetimeFunctionYIntercept{\visitNumber}|$ values are not $<<1$ for high $\laborCost$. This can cause the difference between the $\visitExpenditure{\visitNumber}$ values calculated by EVL and ONPC to be relatively large, noticeably affecting the NPC when $\totalNumberOfVisits$ is also large. When the cost of labor is high, the optimal number of visits tends to be low, meaning that even though the error is present, it is not significant in the range of $\totalNumberOfVisits$ where NPC is minimized.

\subsection{Analysis of NPC Minimization Parameters}
\label{ssec:analysis-of-npc-minimization-parameters}

In this section we examine how the performance of NPC minimization is affected by changes in the parameters described in \sref{ssec:individual-simulation-parameters}. To determine the situations where NPC minimization is most effective under our network model, we will find the parameters that have the biggest impact on the optimal NPC. The parameters considered are the cost $\nodeCost{}$ of node hardware, the cost $\energyCost$ of a Joule, the rate $\powerConsumption$ of energy consumption, the cost $\laborCost$ of labor, the interest rate $\interestRate$, the network's operational lifetime $\operationalLifetime$, and its MTBF $\mtbf$.

\begin{figure}
  \centering
  \includegraphics[width=0.33\textwidth]{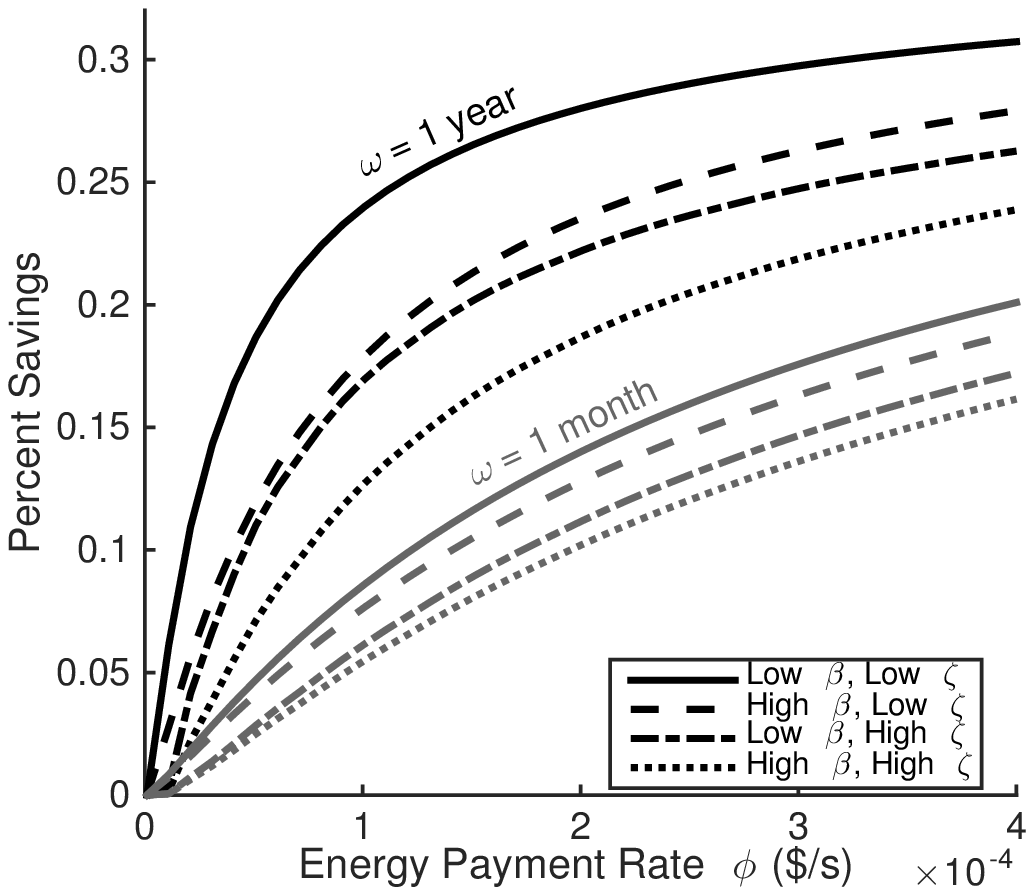}
  \caption{Percent savings of NPC minimization compared to performing a single visit, for various energy payment rates $\energyPaymentRate$ at different node hardware $\nodeCost{}$ costs, different labor $\laborCost$ costs, and different MTBFs $\mtbf$.}
  \label{fig:energy-cost-factors}
\end{figure}

Two parameters that influence the return earned on an investment are $\operationalLifetime$ and $\interestRate$; NPC minimization uses the returns provided by interest rates to reduce future energy expenditures. If either $\operationalLifetime$ or $\interestRate$ is low, the percent savings is less than 5\%, meaning that NPC minimization is ineffective. When $\operationalLifetime$ is low, little time is available to earn interest on the investment. If $\interestRate$ is low, returns will be small even if the money is invested for a number of years.

\fref{fig:energy-cost-factors} shows how different parameters affect the percent savings provided by NPC. The x-axis is the \emph{energy payment rate} $\energyPaymentRate$, which is the cost $\energyCost$ of a Joule multiplied by the rate $\powerConsumption$ at which the network consumes energy, and has units of \$/s. The low and high values for node hardware and labor costs are discussed in \sref{ssec:individual-simulation-parameters}, while the percent savings metric is explained at the beginning of \sref{sec:numerical-results}. 

NPC minimization is most effective when a large portion of the budget is dedicated to energy. In \fref{fig:energy-cost-factors} the percent savings approaches 0 for values of $\energyPaymentRate$ approaching \SI{0}{\$/\second}. Percent savings are higher when $\nodeCost{}$ and $\laborCost$ are low. We can also see that the impact of changes in $\energyPaymentRate$, $\nodeCost{}$, and $\laborCost$ on the percent savings is significantly reduced when the MTBF $\mtbf$ is low. NPC minimization works by deferring expenses to take advantage of interest rates, and in our scenario energy is the only deferrable expenditure, so the greater the portion of the budget dedicated to energy, the greater the percent savings. 

\subsection{Cost Breakdown of NPC Minimization}
\label{ssec:npc-cost-breakdown}

In this section we take a closer look as to how the cost $\nodeCost{}$ of node hardware, cost $\laborCost$ of labor, MTBF $\mtbf$, operational lifetime $\operationalLifetime$, and interest rate $\interestRate$ affect the percent savings and the optimal number of visits. We show that NPC minimization works by balancing decreasing energy costs with increasing labor costs, and study how adjusting each parameter impacts the optimal NPC and number of visits required to achieve this NPC.

\begin{figure}
  \centering
  \includegraphics[width=0.33\textwidth]{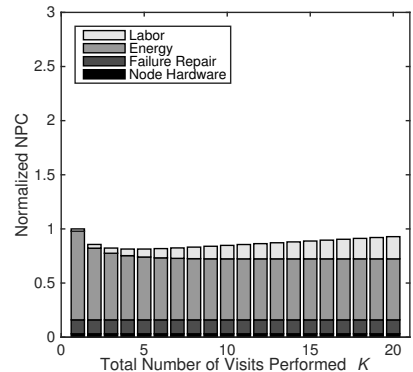}
  \caption{Normalized NPC for different numbers of visits when parameters are set to their default values.}
  \label{fig:default-payments}
\end{figure}

\begin{figure}
  \centering
  \includegraphics[width=0.33\textwidth]{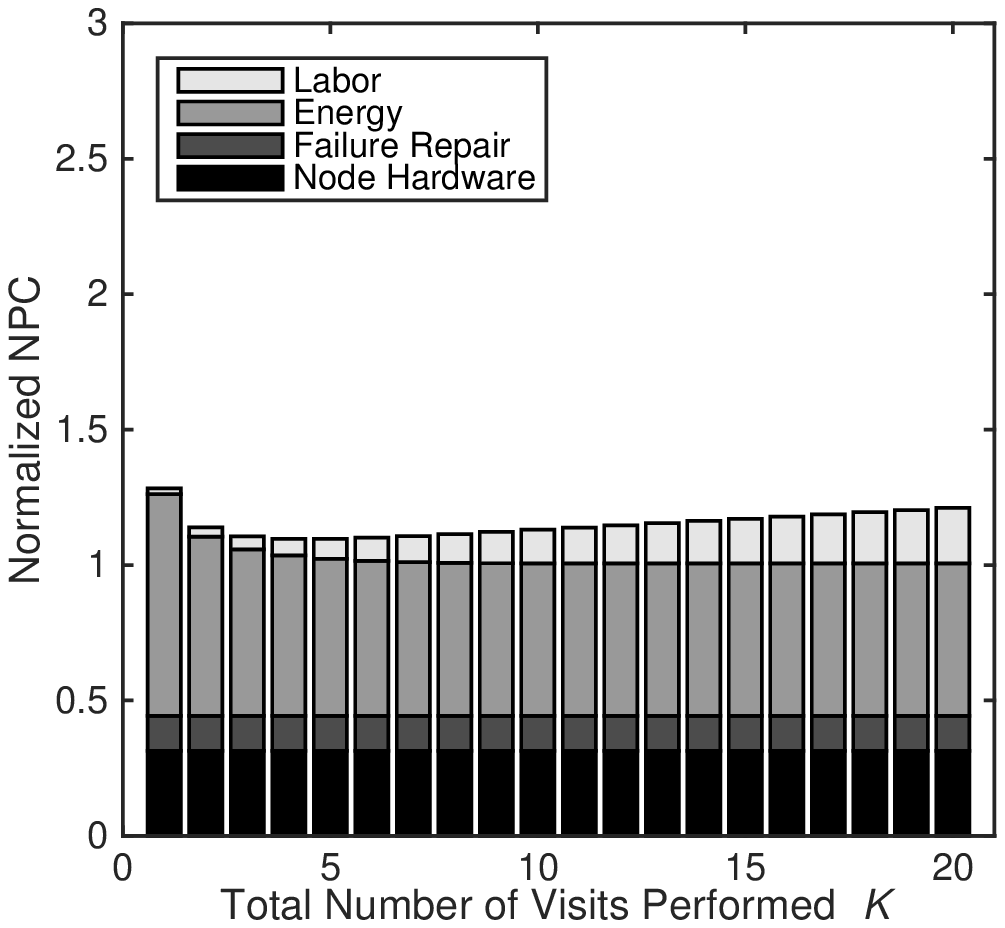}
  \caption{Normalized NPC for different numbers of visits when parameters are set to their default values, except the cost $\nodeCost{}$ of a node which is \$100.}
  \label{fig:high-node-cost-payments}
\end{figure}

\begin{figure}
  \centering
  \includegraphics[width=0.33\textwidth]{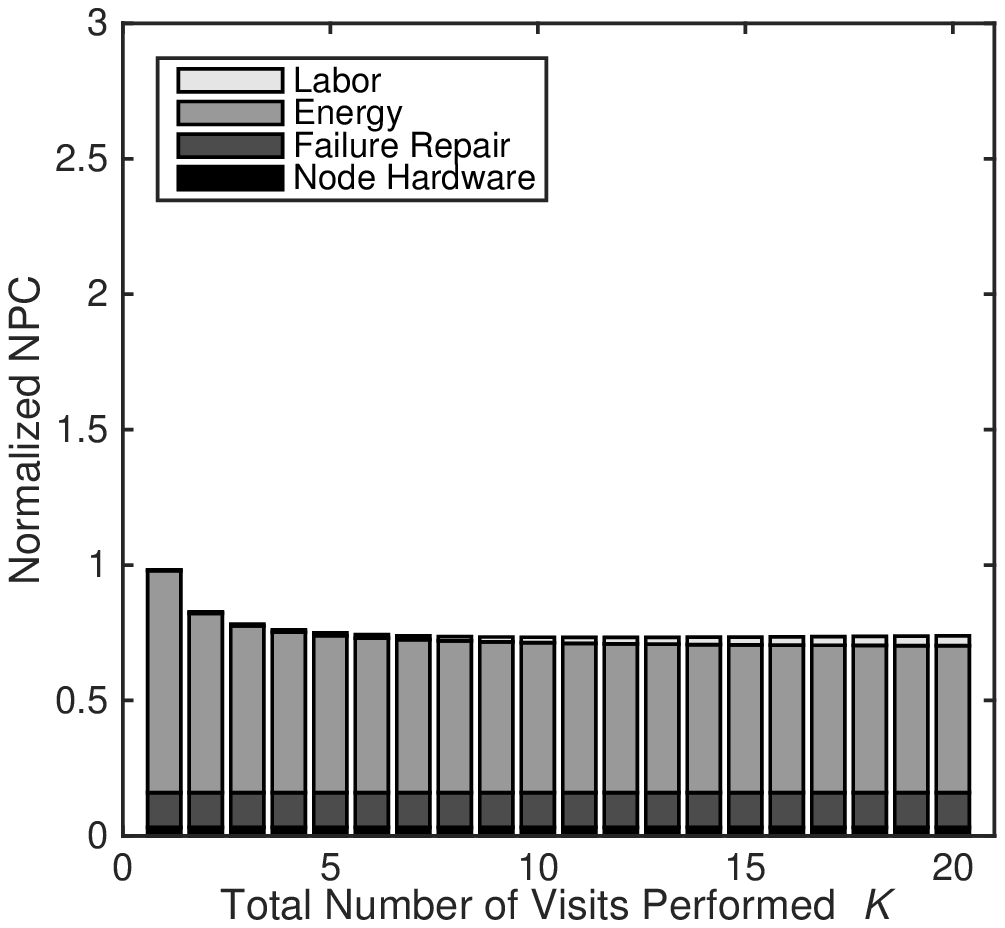}
  \caption{Normalized NPC for different numbers of visits when parameters are set to their default values, except the cost $\laborCost$ of labor which is \SI{140}{\$/visit}.}
  \label{fig:low-maintenance-payments}
\end{figure}

\begin{figure}
  \centering
  \includegraphics[width=0.33\textwidth]{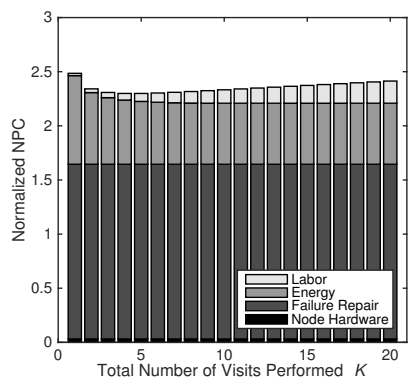}
  \caption{Normalized NPC for different numbers of visits when parameters are set to their default values, except the MTBF $\mtbf$ which is \SI{1}{month}.}
  \label{fig:high-failure-rates}
\end{figure}

Figs. \ref{fig:default-payments}-\ref{fig:high-failure-rates} show the normalized NPC for various numbers of visits. Each bar is segmented to show the expenditures on node hardware, failure repair, energy, and labor for each number of visits being performed. Note that the labor bar segments only include the cost of labor for restoring energy; costs related to labor for failure repair are included in the failure repair bar segments. NPC was normalized to the cost of a network when $\totalNumberOfVisits=1$ for the default parameter values described at the end of \sref{ssec:individual-simulation-parameters}. \fref{fig:default-payments} shows the normalized NPC at the default parameter values. Figs. \ref{fig:high-node-cost-payments}-\ref{fig:high-failure-rates} show the normalized NPC where all parameters are at their default values except one. The adjusted value is given in the figure caption.

\fref{fig:default-payments} shows that NPC minimization balances decreasing energy costs with increasing labor costs. As the number of visits increases, the total cost of energy decreases at a decreasing rate while the total cost of labor appears to increase steadily. As nodes are only purchased once during the first visit, the total cost of nodes does not change. The more a parameter affects the rate that energy costs decrease or labor costs increase, the greater that parameter's effect on the optimal NPC and the number of visits required to achieve it.

\fref{fig:high-node-cost-payments} shows that increasing the cost of node hardware affects the NPC for each value of $\totalNumberOfVisits$ equally. Both the optimal NPC and the cost when performing only one visit increase by the same amount. The reduction in percent savings from increasing $\nodeCost{}$ comes from increasing the NPC when $\totalNumberOfVisits=1$ without changing the difference between it and the optimal NPC. The money saved on energy by performing NPC minimization remains the same as in \fref{fig:default-payments}.

\fref{fig:low-maintenance-payments} shows that decreasing labor costs $\laborCost$ allows a network operator to perform more visits and reduce energy costs further, improving the percent savings. Even performing a large number of visits barely increases the total labor cost. This means that a larger number of visits can be made, and energy costs can be reduced further, before the total cost of labor exceeds the savings from NPC minimization.

\fref{fig:high-failure-rates} shows that increasing the failure rate causes the repair payments to increase uniformly. The number of visits to restore energy does not affect network reliability, explaining why the repair payments increase uniformly regardless of $\totalNumberOfVisits$. A smaller percent of the budget is spent on energy, reducing percent savings without changing the optimal number of visits.

Low interest rates $\interestRate$ and low lifetimes $\operationalLifetime$ rob NPC minimization of its ability to lower energy costs. Returns on invested money at low interest rates and lifetimes are negligible, reducing the savings on future energy costs. When increasing $\totalNumberOfVisits$, the cost of labor immediately exceeds the savings in energy costs, making a single visit optimal when either $\operationalLifetime$ or $\interestRate$ is low.

\section{Conclusion}

We have addressed the problem of minimizing the Net Present Cost (NPC) of operating Wireless Sensor Networks (WSNs) by providing frameworks that determine the number and spacing of visits, as well as the size of their visit expenditures. We provided a general non-linear, non-convex optimization framework for minimizing NPC when the relationship between a visit expenditure and visit lifetime is known. We proposed a framework for maximizing a visit lifetime when given a visit expenditure, and showed that the relationship between visit expenditure and visit lifetime is linear under this framework. We developed a more efficient framework for minimizing NPC that takes advantage of this relationship, and demonstrated that equally spacing visits is a near-optimal strategy under such conditions. 

Compared to making a single visit, NPC minimization can significantly reduce costs by deferring expenditures and using returns on investments and network revenue to reduce future expenditures. It finds the optimal balance between the money saved on energy and the cost of labor required to deliver it to the network. Networks that consume relatively large amounts of energy, networks with long operational lifetimes, and cases where interest rates are high tend to benefit the most from NPC minimization. Large labor costs reduce the number of visits before the total cost of labor exceeds the savings generated by NPC minimization, limiting the opportunities NPC minimization has to lower energy costs. Low lifetimes and interest rates reduce the returns required by NPC minimization.

There are a number of ways in which our NPC minimization framework could be applied by a network operator. In addition to minimizing the overall cost of a network, the network operator can determine the cost of deploying the network, the optimal number of times to visit the network to perform maintenance, the optimal spacing between visits, and the optimal visit expenditures. This information is sufficient for the network operator to predict the CAPEX and OPEX of the network. The network operator can also predict how each visit expenditure will be divided between node hardware, energy, and labor costs. A maintenance schedule that not only predicts when to visit the network to perform maintenance, but also the node hardware, energy, and labor resources required, can therefore be created with the NPC minimization framework.

\appendices

\section*{Acknowledgment}

This work was supported by NSERC, DAAD RISE, and the Bell Labs internship program. The authors would like to thank Vinay Suryaprakash (Bell Labs, Alcatel-Lucent, Germany) for his valuable comments.

\ifCLASSOPTIONcaptionsoff
  \newpage
\fi

\bibliographystyle{IEEEtran}
\bibliography{overview,energy,maintenance,node_placement,optimal_energy,optimizers,geosteiner,hardware,cellular_deployment,cost_minimization,miscellaneous}

\newcommand{\biospace}{-0.33in}
\vspace{\biospace}
\begin{IEEEbiography}
[{\includegraphics[width=1in,height=1.25in,clip,keepaspectratio]{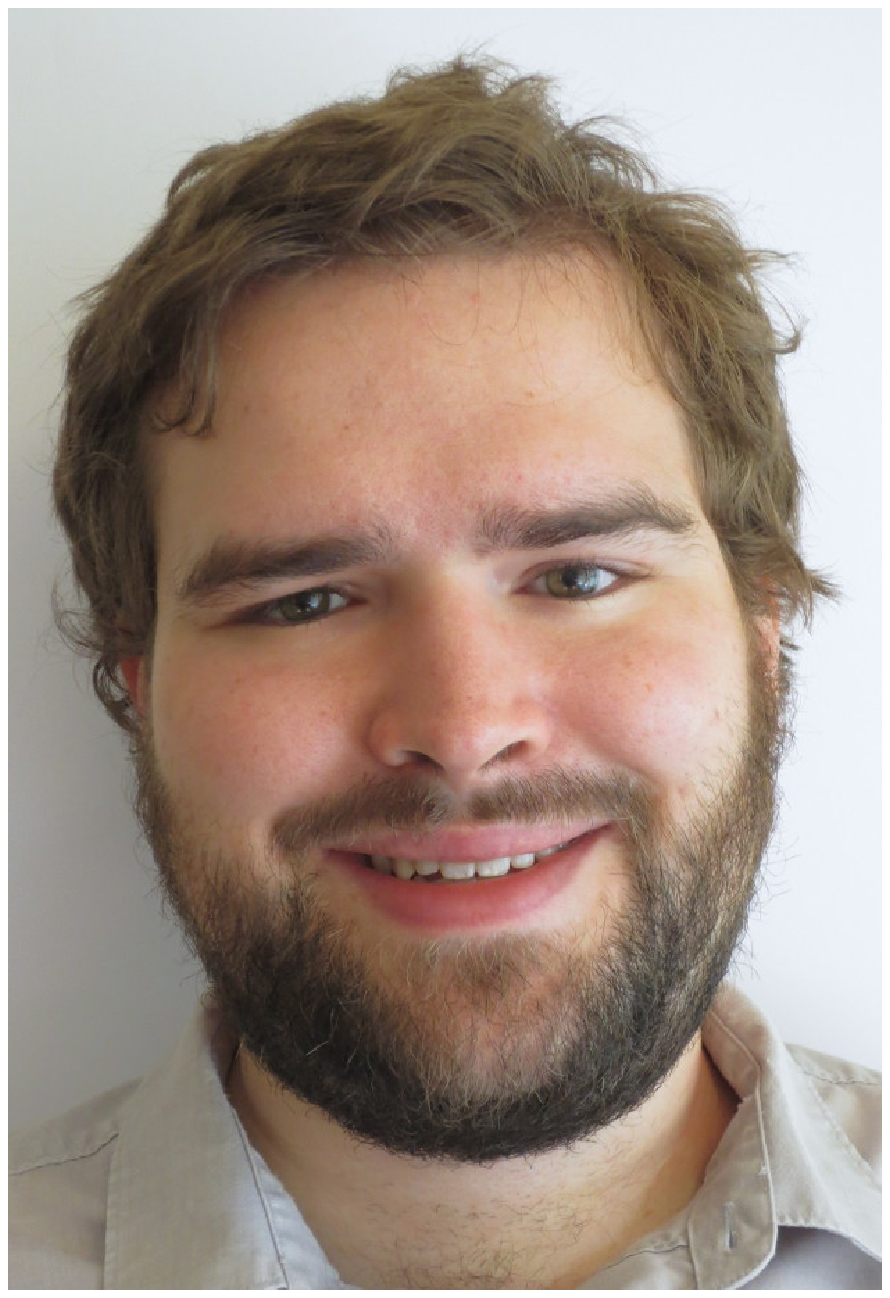}}]
{Kevin Dorling}
(S'14) received a B.S. degree in Computer Engineering with distinction from the University of Calgary, Canada, in 2010. He is currently pursuing a Ph.D. in Electrical Engineering at the University of Calgary.

In 2011 and 2013, he was an intern at Bell Labs, Alcatel-Lucent in Stuttgart, Germany, working on deployment techniques for wireless sensor networks. In 2009, he was an intern at CDL Systems in Calgary, Canada, developing control station software for unmanned vehicles. His research interests include wireless sensor networks, operations research, and communication systems.
\end{IEEEbiography}
\vspace{\biospace}
\begin{IEEEbiography}
[{\includegraphics[width=1in,height=1.25in,clip,keepaspectratio]{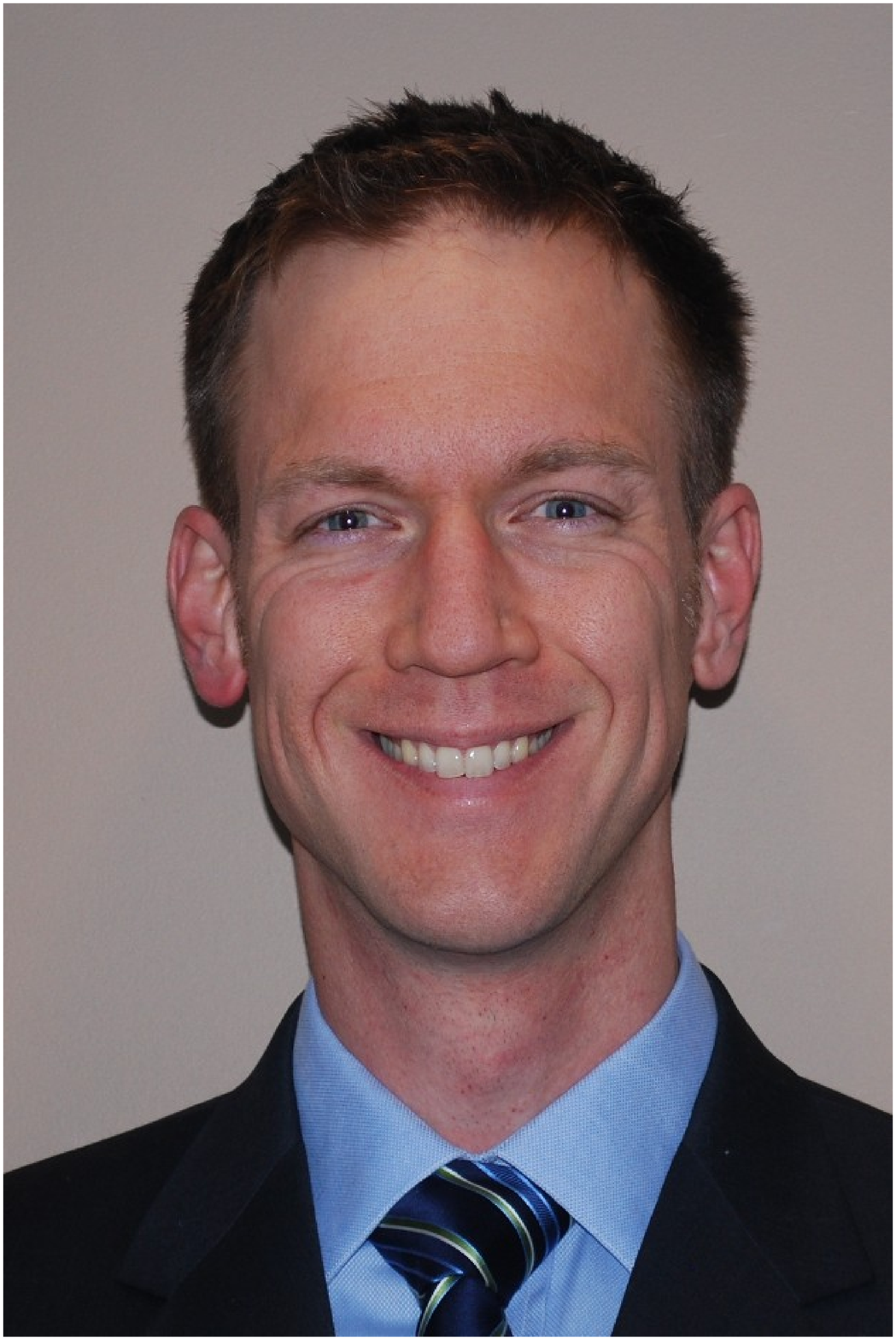}}]
{Geoffrey Messier}
(S'91 - M'98) received his B.S. in Electrical
Engineering and B.S. in Computer Science degrees from the University of
Saskatchewan, Canada with great distinction in 1996.  He received his
M.Sc. in Electrical Engineering from the University of Calgary, Canada
in 1998 and his Ph.D. degree in Electrical and Computer Engineering from
the University of Alberta, Canada in 2004.

From 1998 to 2004, he was employed in the Nortel Networks CDMA Base
Station Hardware Systems Design group in Calgary, Canada.  At Nortel
Networks, he was responsible for radio channel propagation measurements
and simulating the physical layer performance of high speed CDMA and
multiple antenna wireless systems.  Currently, Dr. Messier is a
Professor in the University of Calgary Department of
Electrical and Computer Engineering.  His research interests include
data networks, physical layer communications and communications
channel propagation measurements.
\end{IEEEbiography}
\vspace{\biospace}
\begin{IEEEbiography}
[{\includegraphics[width=1in,height=1.25in,clip,keepaspectratio]{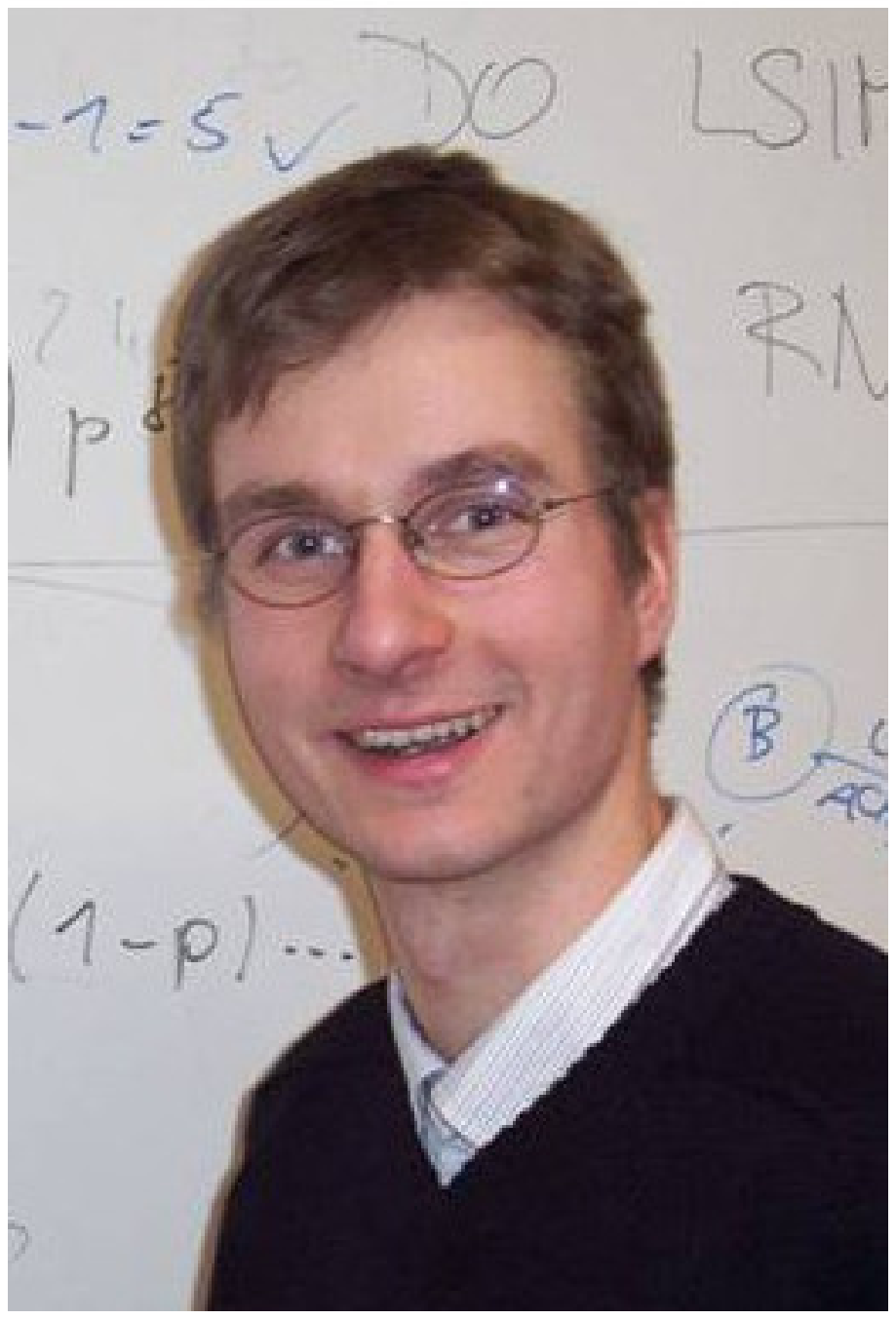}}]
{Stefan Valentin}
(S'07, M'09) received an M.A. in EE with excellence from the Technical University of Berlin, Germany in 2004 and a Dr. rer. nat. in CS with summa cum laude from the University of Paderborn, Germany in 2010. In the same year, he joined Bell Labs, Stuttgart, Germany as a Member of Technical Staff, where he worked on wireless resource allocation algorithms for 4G and 5G. In 2015, he moved to Paris, France where he is leading the Context-Aware Optimization team at Huawei's new Mathematical and Algorithmic Sciences Lab. Dr Valentin's research interests are network information theory, cooperative relaying, and wireless resource allocation. Dr Valentin received the ACM SIMUTools Best Paper Award in 2008, the Bell Labs Award for Exceptional Achievements in 2011, the Klaus Tschira Award for Comprehensible Science in 2011, and the Bell Labs Award of Excellence in 2013. In 2015, he co-received Huawei's Award of Excellence, and the Fred W. Ellersick Prize of the IEEE Communications Society.
\end{IEEEbiography}
\vspace{\biospace}
\begin{IEEEbiography}
[{\includegraphics[width=1in,height=1.25in,clip,keepaspectratio]{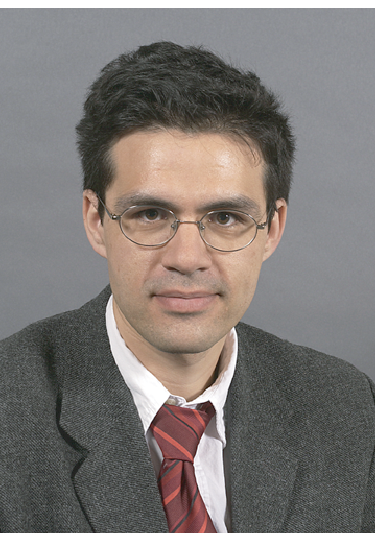}}]
{Sebastian Magierowski} received his Ph.D. degree in Electrical Engineering from the University of Toronto in 2004.  From 2004 to 2012 he served as Assistant/Associate Professor in the Department of Electrical and Computer Engineering at the University of Calgary after which he joined the faculty of the Department of Electrical Engineering and Computer Science in the Lassonde School of Engineering at York University, Toronto, Canada.  

As part of his industrial experience (Nortel Networks, PMC-Sierra, Protolinx Corp.) Dr. Magierowski has worked on CMOS device modeling, high-speed mixed-signal IC design, and data networks.  His research interests include analog/digital CMOS circuit design, communication systems and biomedical instrumentation.
\end{IEEEbiography}

\end{document}